\UseRawInputEncoding

\documentclass[galaxies,article,accept,moreauthor,pdftex,12pt,a4paper]{mdpi}
\usepackage{soul}
\usepackage{upgreek}
\usepackage{enumitem}
\usepackage{paralist}
\usepackage{subfigure}
\usepackage{booktabs}
\usepackage{multirow}
\usepackage{longtable}
\usepackage{microtype}
\usepackage{xcolor}
\newcommand\myurl[1]{\changeurlcolor{black}\url{#1}\changeurlcolor{blue}}

\usepackage{subfigure}
\makeatletter
\renewcommand{\@thesubfigure}{\normalsize(\textbf{\alph{subfigure}})}
\makeatother
\newcommand{\Mov}[1]{{\color{black}{#1}}}

\firstpage{1}
\articlenumber{??}
\doinum{10.3390/galaxies???}
\pubvolume{??}
\pubyear{2021?}
\copyrightyear{2021?}
\externaleditor{Academic Editors: Jose Gaite and Antonaldo Diaferio?}
\history{Received: {? June 2021}; Accepted: {???}; Published: 
{???}}
\pdfoutput=1

 \theoremstyle{mdpi}
 \newcounter{thm}
 \newcounter{ex}
 \newcounter{re}
 \theoremstyle{mdpidefinition}

\def\aap{Astron. Astrophys.}
\def\aj{Astro. J.}
\def\apss{ Astrophys.  Space Sci. }
\def\apj{Astrophys. J.}
\def\apjl{Astrophys. J. Lett.}
\def\apjs{Astrophys. J. Supp.}
\def\jcap{J. Cosmo. Astrop. Phys}
\def\mnras{Mon. Not. Roy. Astro. Soc.}
\def\na{New Astron.}
\def\nar{New Astron. Rev.}
\def\nat{Nature}

\def\prd{Phys. Rev. D}
\def\physrep{Phys. Rept.}
\def\araa{Ann. Rev. Astron.  Astrophys.}

\usepackage{amsmath}

\def\nvphantom{\v@true\h@false\nph@nt}
\def\nhphantom{\v@false\h@true\nph@nt}
\def\nphantom{\v@true\h@true\nph@nt}
\def\nph@nt{\ifmmode\def\next{\mathpalette\nmathph@nt}%
  \else\let\next\nmakeph@nt\fi\next}
\def\nmakeph@nt#1{\setbox\z@\hbox{#1}\nfinph@nt}
\def\nmathph@nt#1#2{\setbox\z@\hbox{$\m@th#1{#2}$}\nfinph@nt}
\def\nfinph@nt{\setbox\tw@\null
  \ifv@ \ht\tw@\ht\z@ \dp\tw@\dp\z@\fi
  \ifh@ \wd\tw@-\wd\z@\fi \box\tw@}

\Title{Review of solutions to the Cusp-core problem of the $\Lambda$CDM Model}

\Author{Antonino Del Popolo $^{1,2,3,
}$ and Morgan Le Delliou $^{4,5,6
}$*}
\AuthorNames{Antonino Del Popolo and Morgan Le Delliou}

\address{$^{1}$ \quad Dipartimento di Fisica e Astronomia, University of Catania
, Viale Andrea Doria 6, {95125}
Catania, Italy
 \quad  \\
$^{2}$  \quad INFN Sezione di Catania,  Via S. Sofia 64, I-95123 Catania, Italy\quad  \\
$^{3}$ \quad International Institute of Physics, Universidade Federal do Rio Grande do Norte,
59012-970 Natal, Brazil\quad  \\
$^{4}$ \quad Institute of Theoretical Physics, School of Physical Science and Technology, Lanzhou University,
No.222, South Tianshui Road, Lanzhou, Gansu 730000, P R China
 \\
$^{5}$ \quad Instituto de Astrof\'isica e Ci\^encias do Espa\c co, Universidade de Lisboa,
Faculdade de Ci\^encias, Ed.~C8, Campo Grande, 1769-016 https://www.overleaf.com/6626887373nrqgfwmnbsdr
￼
Lisboa, Portugal\\
$^6$ \quad Lanzhou Center for Theoretical Physics, Key Laboratory of Theoretical Physics of Gansu Province, Lanzhou University, Lanzhou, Gansu 730000, China
}

\corres{Correspondence: (delliou@lzu.edu.cn,)morgan.ledelliou.ift@gmail.com
}

\firstnote{adelpopolo@oact.inaf.it
}
\secondnote{(delliou@lzu.edu.cn,)morgan.ledelliou.ift@gmail.com
}

\abstract{
This review aims at proposing to the field an overview of the Cusp-core problem, including a discussion of its advocated solutions, assessing how each can satisfactorily provide a description of central densities.
Whether the Cusp-core problem reflects our insufficient grasp on the nature of dark matter, of gravity, on the impact of baryonic interactions with dark matter at those scales, as included in semi-analytical models or fully numerical codes, the solutions to it can point either to the need for a paradigm change in cosmology, or to to our lack of success in ironing out the finer details of the $\Lambda$CDM paradigm. 
}

\begin{document}



\section{Introduction}

Notwithstanding the many successes of the concordance cosmology model, also known as the $\Lambda$CDM model, fundamental issues remain open.
Most of the cosmology community considers the $\Lambda$CDM paradigm to accurately describe the Universe's large scale structure formation and evolution, its early state and the proportions of its content in matter and energy \cite{spergel03,komatsu11,dp07,dp13,dp14a}. This view has been reinforced with each new \Mov{discovery}
, such as the kinetic Sunyaev-Zel'Dovich (SZ) effect, the Cosmic Microwave Background (CMB) B-mode polarisation \cite{hanson}, or its lensing \cite{smith,das},...
Among its most famous resisting issues, the questions of the nature, ``fine tuning'' and ``cosmic coincidence'' problems of the cosmological constant $\Lambda$ remain open \protect\cite{weinberg_co,ADP}. 

The ``cosmological constant fine tuning problem'' emerges from the \Mov{large discrepancy }
between the observed value of $\Lambda$ and the huge values predicted by quantum field theories for the present quantum vacuum energy, the latter exceeding the former by more than 100 orders of magnitude \protect\citep[][]{weinberg_co,martin_j,ADP}: the cosmological observation upper bounds are limited to $\rho_{\Lambda} \simeq 10^{-47} ~\rm GeV^4$, while the naive theoretical expectations obtain $\rho_{\Lambda} \simeq 10^{71} ~\rm GeV^4$. This is the most extreme fine tuning problem known to physics. As $\Lambda$ is constant, the concordance model also presents a strong fine tuning of the dark energy (DE) initial conditions. 
The ``cosmic coincidence problem'' questions the relative coincidence of  the orders of magnitude at present of dark matter (DM) and DE energy densities \protect\citep[][]{sivanandam}.

Further issues of the $\Lambda$CDM paradigm revolve around observations of the CMB or of smaller scales structures. 
Unexplained statistical anomalies in the CMB large-angle fluctuations contradict the $\Lambda$CDM CMB multipoles statistical independence, which  reflects the assumption that the content of our Universe should realise a Gaussian and statistically isotropic random field. These anomalies comprise detection in the CMB of a power hemispherical asymmetry \cite{eriksen,hansen,jaf,hot,planck1,akrami}, of a cold spot, marking an underdense region, \cite{cruz,cruz1,cruz2}, and of a quadrupole-octupole alignment \cite{schwa,copi,copi1,copi2,copi3}. Such anomalies could either mark new physics, be caused by statistical effects \cite{bennet}, or reflect unknown systematics.

Furthermore, puzzling tensions have been found between the cosmological fluctuations $\sigma_8$ parameters measured by cluster number counts and weak lensing, compared with the 2013 Planck results \cite{planck}, as well as between the 2013 Planck $H_0$ Hubble parameter, compared with SNIa data. Those tensions have persisted between the Planck 2015 data and the $\sigma_8$ growth rate \cite{maca}, as well as with the CFHTLenS weak lensing \cite{raveri}. 

In addition to those CMB issues, the description of small scale structures in the $\Lambda$CDM model also raise problems \citep[e.g.][]{moore94,moore1,ostrik,boyl,boyl1,oh}. They count the ``missing satellite problem'' \cite[MSP][]{Klypin:1999uc,moore1}, the ``Too Big To Fail'' problem \cite[TBTF][]{boyl,boyl1}, the angular momentum catastrophe \cite[AMC][]{vanbuswa,Cardone:2009jr}, the satellites planes problem \cite{pawl}, and the baryonic Tully-Fisher relation reproduction problem \cite{mcgaugh}. Among those small-scale problems, the so called Cusp-core problem \cite[CC][]{moore94,flores} retains a central role. The CC coins the gap between the flat central slopes of density profiles in dwarf galaxies (i.e. dwarfs), Low Surface Brightness (LSBs), and Irregulars galaxies on one hand, while they all are DM dominated, and on the other hand, the cuspy central slopes of dissipationless N-body simulations density profile predictions \cite{nfw,navarro10,Saburova:2014opa}.

Besides the aforementioned issues, further problems of the $\Lambda$CDM paradigm \cite{Kroupa:2004pt,Kroupa:2010hf,Kroupa:2012qj,Kroupa:2013yd,Kroupa:2014ria} are left out of this review, as we focus here on the CC problem. Note however that the formation of cusps and issues of satellites in $\Lambda$CDM are tightly connected, as shown through the mechanism in some unified models \cite{zolo,brooks,dpet14} that can turn cusps into cores as well as alleviate the MSP, since the shape of the parent halo determines strongly its tidal effect on satellites \citep[e.g.][]{Mashchenko2006,Mashchenko2008,pen10}, with cored profiles even capable of stripping the satellite out of existence \cite{pen10}.

We organise the discussion as follows:
starting with a short exposition of the initial discussions on the CC problem, Sec.~\ref{sec:ccp} will summarise its issues, while early solutions involving baryons, DM matter type changes or gravity modifications will be presented in Sec.~\ref{sec:earlySol}. A review of the baryonic solutions to SPH simulations collapse models through gas heating feedback effects from dynamical friction of baryonic clumps will be discussed in Sec.~\ref{sub:BsolCC}, including a discussion in Sec.~\ref{sub:SNsolCC}, where the supernovae explosions~\cite{SomEtal99} solutions of core heating will be presented, 
a Sec.~\ref{sub:DFclumps} that will discuss both baryonic approaches and a unified baryonic solution \citep[][]{dpet14} and a presentation of the resulting mass dependent density profile in Sec.~\ref{sub:MassDepDen}. Finally, Sec.~\ref{sec:CosmoSol} will review and discuss solutions modifying the DM nature, before to conclude in Sec.~\ref{sec:concl}.

\section{Exposition of the Cusp/Core Problem }\label{sec:ccp}

\begin{figure}
\centering
\includegraphics[width=7cm]{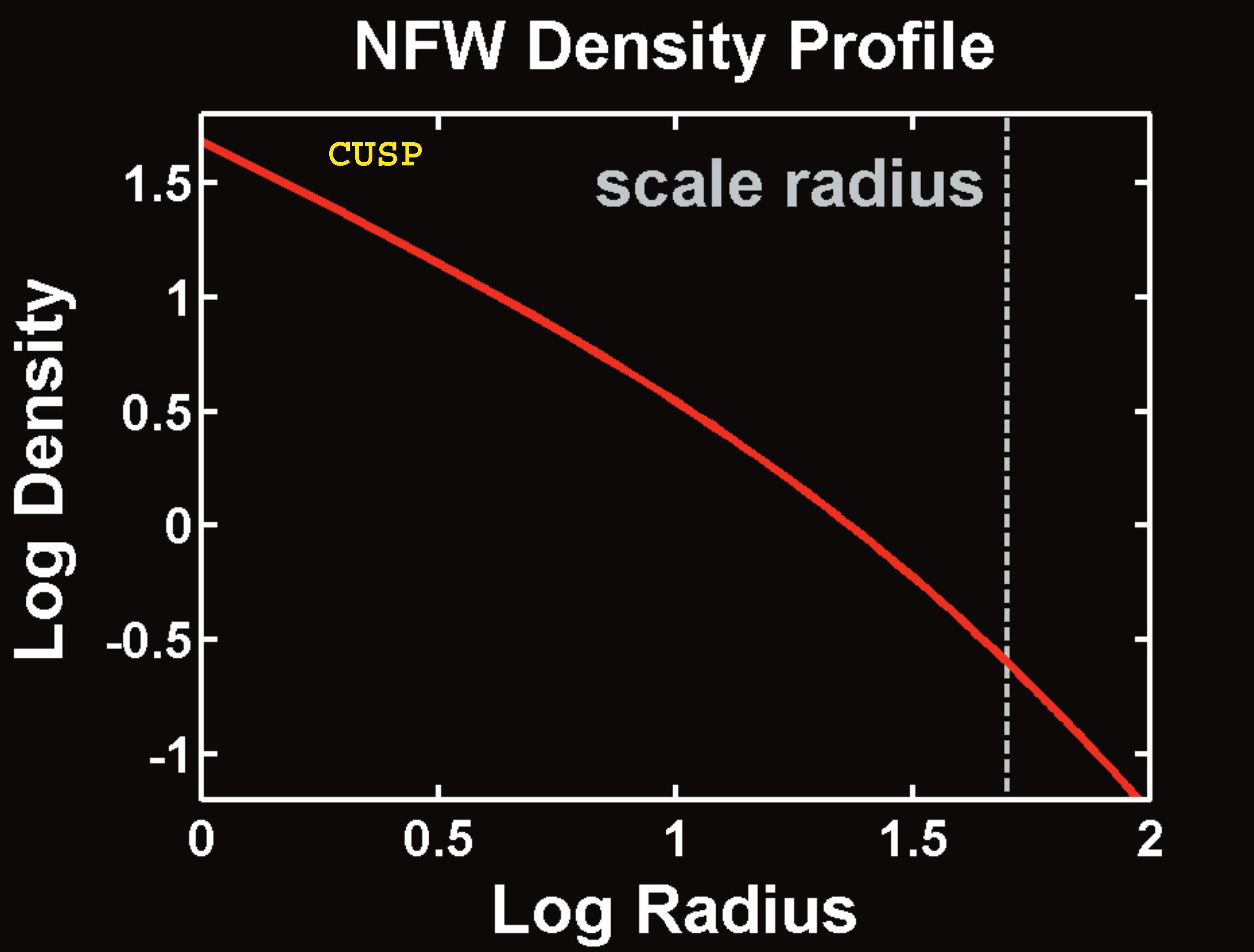}
\includegraphics[width=7cm]{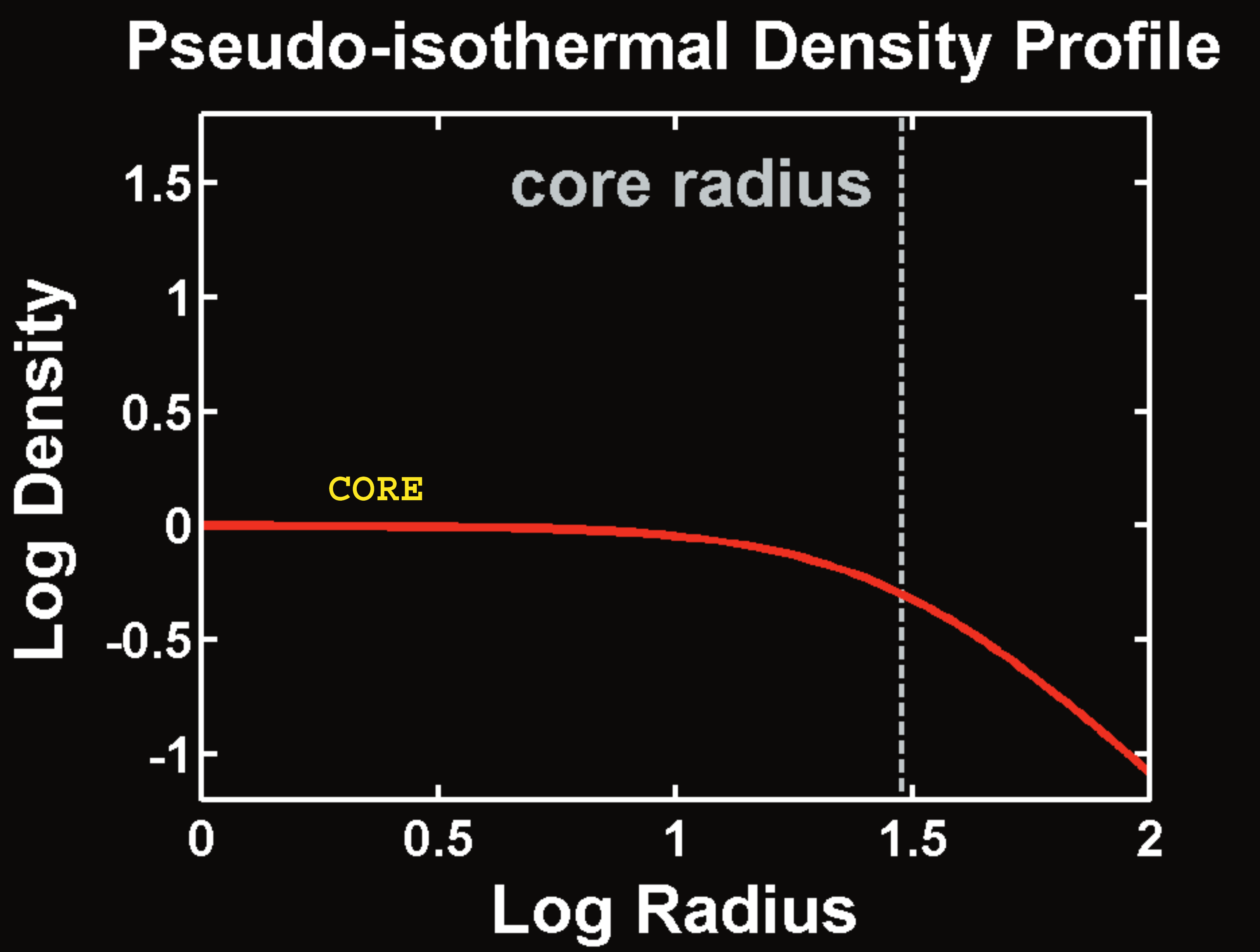}
\caption{Illustration of the CC problem. The cuspy NFW profile is displayed on the left panel, with the dotted vertical line representing the profile's scale radius. The~right panel exemplifies the cored profile, with the dotted vertical line marking the profile's core radius.}
\label{fig:spectra}
\end{figure}

The CC problem emerged after Flores \& Primack \protect\cite{flores} and Moore \protect\cite{moore94} showed DDO galaxies' rotation curves could be better fitted with cored (or pseudo-)isothermal density profiles, ruling out cuspy profiles. As dissipationless CDM model simulations persisted to produce cuspy profiles (see~Fig.~\ref{fig:spectra}), this was coined as the CC problem.

\begin{figure}[H]
\centering
\includegraphics[width=14cm]{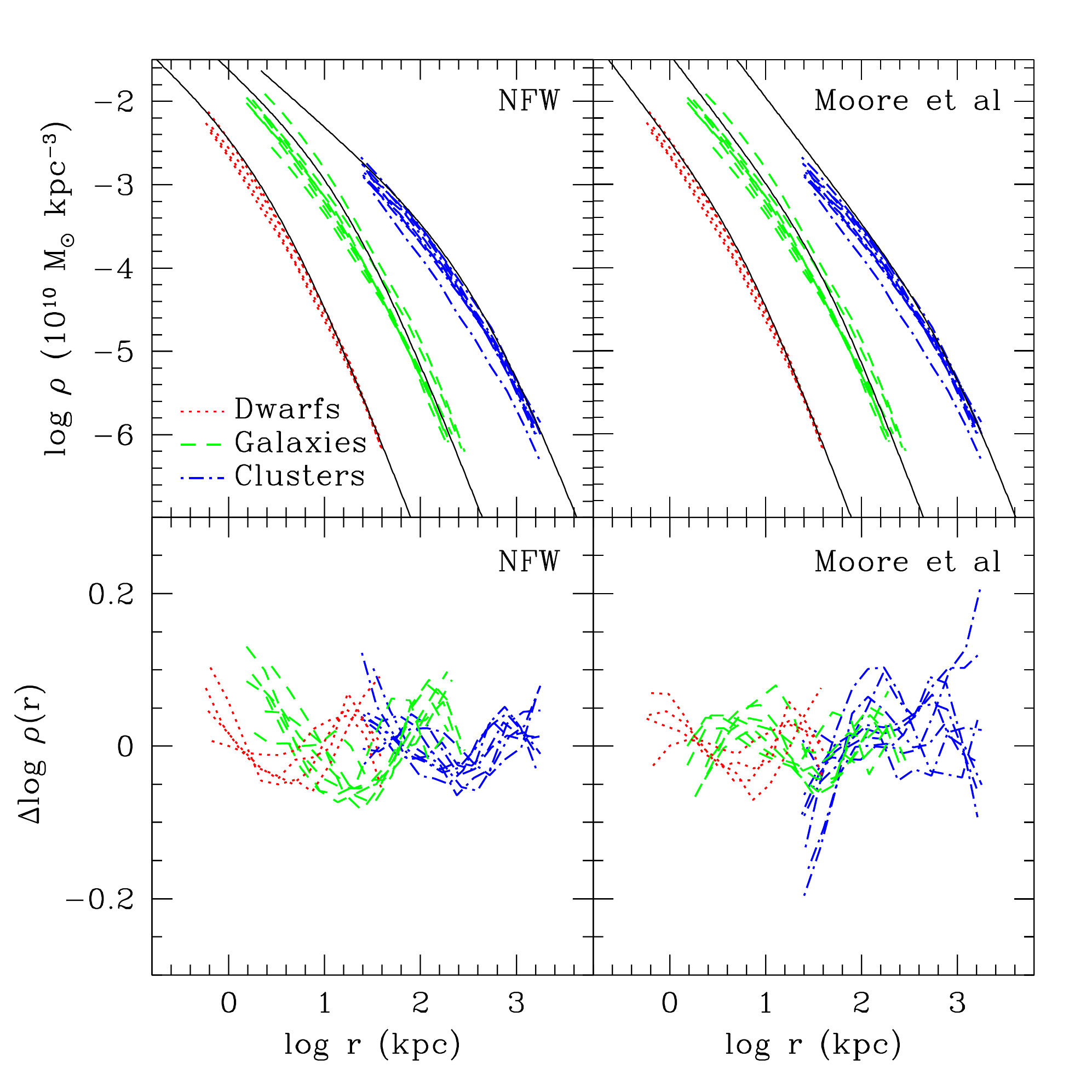}
\vspace{-6pt}
\caption{Comparison of density profiles with simulations.  The NFW profile (solid~lines) is compared in the top left panel with simulations from Ref.~\cite{navarro04}. Note that the Einasto profile fits also well those simulations \cite{navarro10}. The same comparison for the Moore profile is displayed in the top right panel, and each panel represents simulations of dwarfs galaxies (red line), medium size galaxies (green line), and galaxy clusters (blue line). The bottom left and right panels focus on the residuals between the corresponding theoretical profiles and the compared simulations \citep[this figure is reproduced from][Fig.~1]{navarro04}.
}
\label{fig:NFWmoore}
\end{figure}

Indeed, such cuspy DM profile was obtained in the dissipationless N-body simulations of Navarro, Frenk, \& White \protect\cite{nfw,nfw1}, yielding inner density $\rho \propto r^\alpha$, for the power index $\alpha=-1$. Moreover, they found such profile (thus coined NFW profile) to emerge universally, thus independently of cosmology and scale, from dissipationless simulations. This universality was challenged by the obtaining of a steeper profile at $\alpha=-1.5$ \cite{Moore1998,Fukushige2001}, or even by studies observing dependence of the inner slope on the nature of the object and/or on its mass \citep{Jing2000,Ricotti2003,Ricotti2004,Ricotti2007,DelPopolo2010,Cardone2011b,DelPopolo2011,DelPopolo2013d,
DiCintio2014}. Finally (see Fig.~\ref{fig:NFWmoore}), the so-called Einasto profile, yielding a central flattening down to $\simeq\!\!-0.8$ \citep{Stadel2009}, tend to better fit more recent N-body dissipationless simulations \citep[]{Gao2008}. Nevertheless, the CC problem persists, and has been strongly debated for two decades, as even this smallest predicted inner slope from dissipationless N-body simulations exceeds slopes outputted from SPH simulations \citep{Governato2010,Governato2012}, semi-analytical models
\citep{DelPopolo2009,Cardone2012,DelPopolo2012a,DelPopolo2012b, dpc2012,DelPopolo2014a,Popolo:1999im} or even observations \citep{Burkert1995,deBlok2003,Swaters2003,KuziodeNaray2011,Oh2011a,oh}.

Historically, the first contradicting results originated from \Mov{dwarf and} LSB observation in HI: their rotation curves (RCs) did not match the NFW halo \Mov{\cite{Hernandez:1998hf,mcgaugh_deblok,cote}}. Fueling the debate, steep profiles could not be excluded for several objects studied by van den Bosch \& Swaters \cite{vdbswater}. However evidence against steep profiles further mounted through H$\alpha$ data, in de Blok \& Bosma \cite{deblok_bosma}, in the inner slopes $\alpha = -0.2 \pm 0.2$ measured by de Blok et al. \cite{deBlok2003}, and $\alpha= -0.22 \pm 0.08$ obtained by Spekkens et al. \cite{spekkens}, although contradicted by RCs agreeing with cusply profiles produced by Hayashi et al.  \cite{hayashi}. Nevertheless, flat profiles were usually in agreement with high resolution observations \citep[][etc.]{blaise,kuzio08,kuzio09,weldrake,trachte}.

\begin{figure}
\begin{center}
\phantom{e}\hspace{-1.5cm}
\includegraphics[width=15cm]{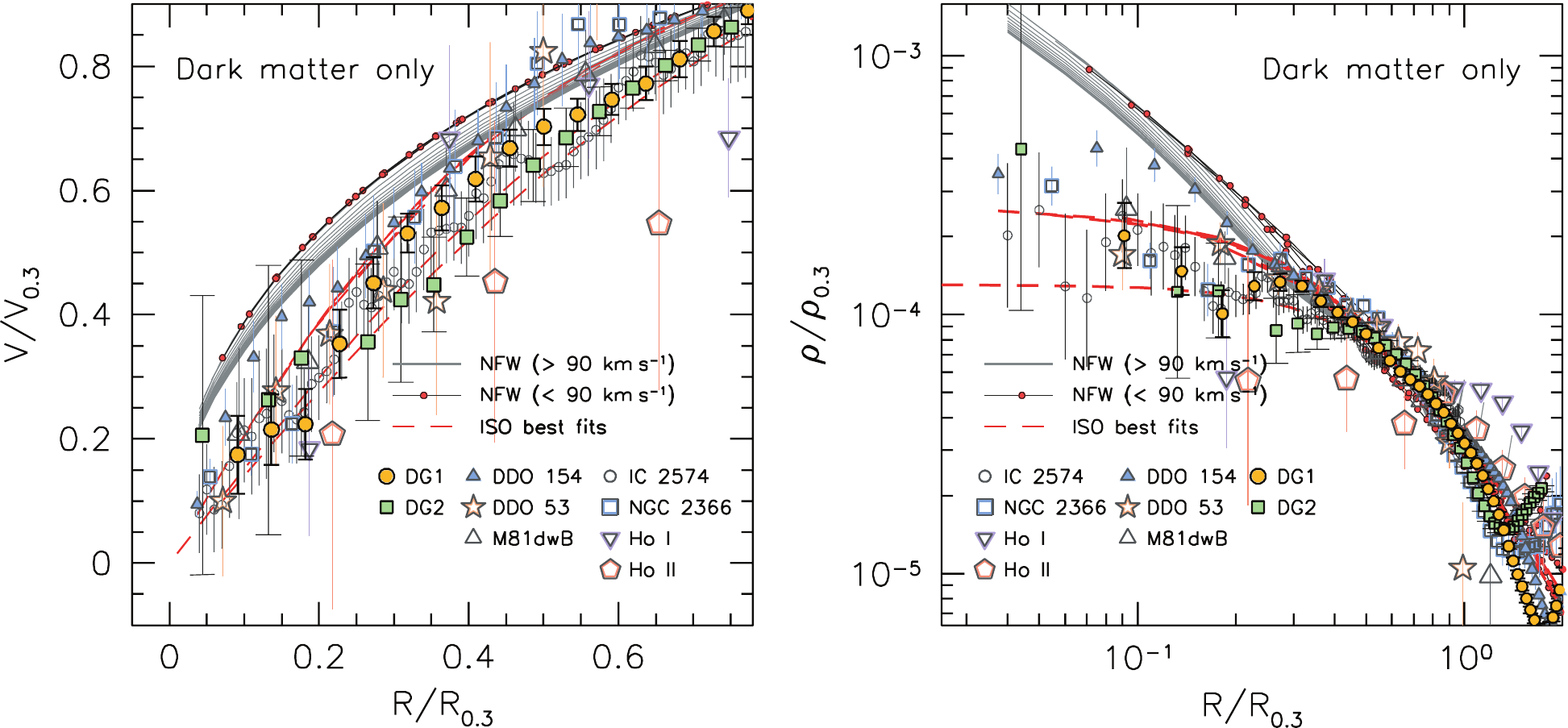}\hspace{2cm}
\end{center}
\vspace{-6pt}
\caption[]{

Confrontations between RCs (left panel) or density profiles (right panel) from the pseudo-isothermal (ISO, red dashed line) and the NFW (solid lines) profiles, to thoses from (\textbf{a}) the \cite{Governato2010} simulated galaxies marked DG1 and DG2, and (\textbf{b}) 7 dwarf galaxies selected from THINGS. The small red dots emphasize the NFW RCs for $V_{200}\in$ [10--90] km/s. RCs are normalised to the rotation velocity $V$ at the distance where $\frac{d \log V}{d \log R}=0.3$, noted $R_{0.3}$ \citep[this figure is reproduced from][Fig.~6]{oh}
. 
}\label{fig:RCdenNFWiso}
\end{figure}

Understanding of the construction of RCs was refined with their decomposition for some spiral galaxies into their components in terms of stars, gas and DM, fitted with various models in Gentile et al.~\protect\cite{gentile,gentile1,gentile2}. The results favoured constant density core profiles over cuspy models, in the same way as the THINGS (The HI Nearby Galaxy Survey) 7 dwarf galaxies processed by Oh et al. \protect\cite{oh}. Fig.~\ref{fig:RCdenNFWiso} compares the RCs of the latter, among others, with those of NFW and pseudo-isothermal (ISO) profiles, clearly showing the ISO fits them best. The LITTLE THINGS survey also produced similar results \cite{Oh2015}.

However, as the previous results emanate mainly from low brightness objects such as dwarfs or LSBs, while large galaxies or high-surface brightness objects  inner structure determination is more complex, general statements on inner density profiles of all galaxies, in particular on their cored or cuspy nature, are far from settled. Indeed, while the high-surface brightness galaxies studied by \cite{Spano2008} present cored centres, their analysis by \citep[e.g.,][]{Simon2005,deBlok2008,dpc2012,DelPopolo2013d,Martinsson2013} disagree. Other surveys contain a mix of cored and cuspy objects. In particular, in the THING sample, galaxies with low luminosity, $M_B>-19$, tend to follow a cored ISO profile, while luminous galaxies, $M_B<-19$,  are equally well fitted by cored or cuspy profiles.

\begin{figure}[H]
\hspace{-0.5cm}
\centering
\includegraphics[width=15cm]{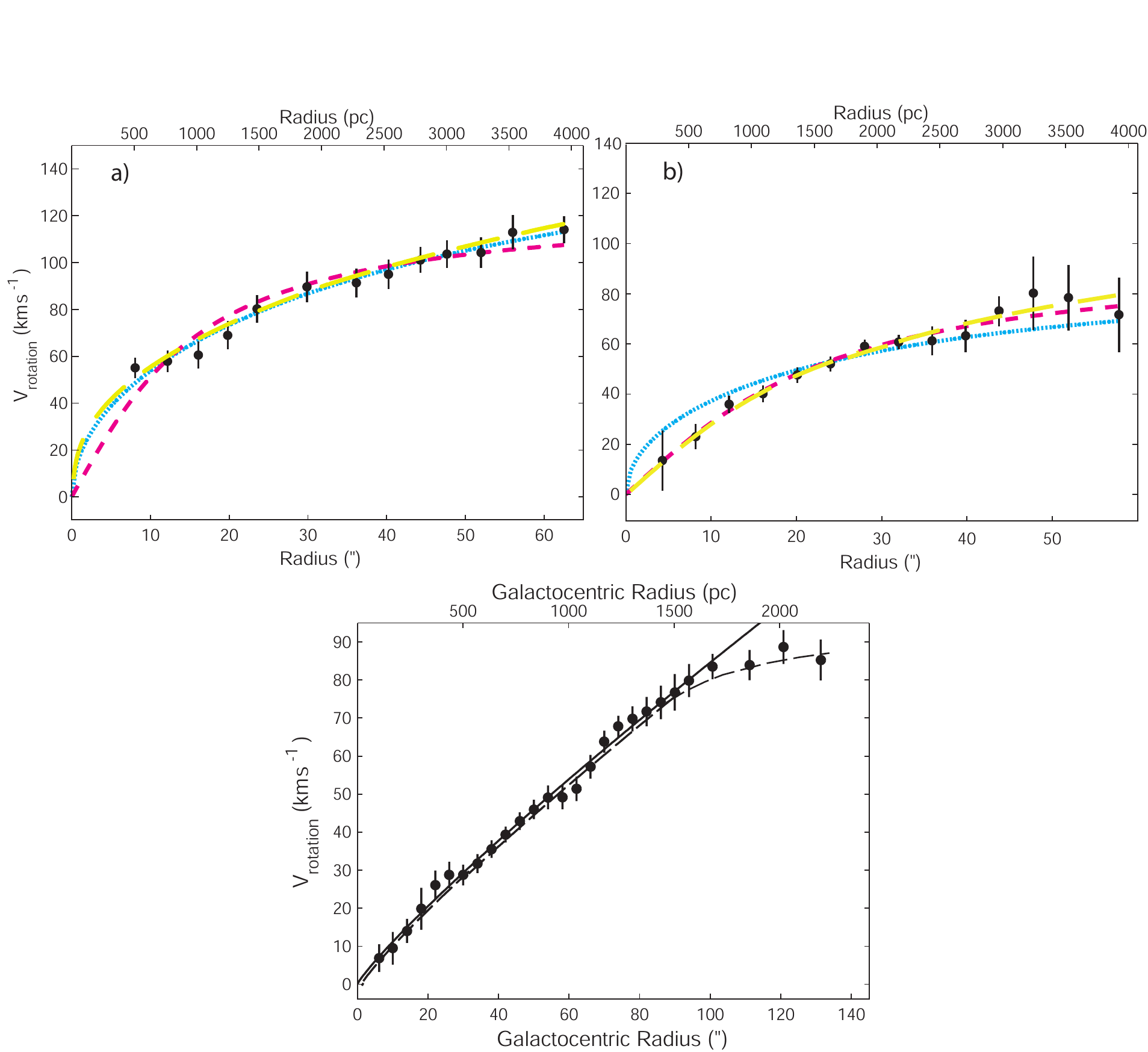}
\caption[]{Presentation of the DM halo RCs  (error-barred, black dots) for the galaxies NGC5963, NGC5949 and NGC2976, provided in Ref.~\cite{Simon2005}. In the two top panels showing resp. NGC5963, on top left, and NGC5949, on top right, are represented NFW profile fitted RCs (dotted cyan lines), pseudo-isothermal profile fitted RCs  (ISO, short-dashed magenta lines) and the RCs from Ref.~\cite{DelPopolo2012a}'s model (yellow dashed lines). The bottom panel, displaying NGC2976, compares its observations with a density power-law fit model (slope $\simeq 0.01$, solid line) and with  Ref.~\cite{DelPopolo2012a}'s model (dashed line; this figure is adapted from Ref.~\cite[Figs.~4 and 5]{DelPopolo2012a}).

}\label{fig:RCobs1}
\end{figure}

Such confused picture is reinforced by the lack of systematically flat slopes in dwarf galaxies: in \mbox{Simon et al. \protect\cite{Simon2005}}, a large scatter in the inner slopes of the low mass spirals NGC2976, NGC4605, NGC5949, NGC5963, and NGC6689, was obtained, with a cored profile $\alpha \simeq \Mov{-}0.01$ for NGC2976, a cuspy slope $\alpha \simeq \Mov{-}1.28$ for NGC5963, while $\alpha \simeq \Mov{-}0.88$ for NGC4605, $\alpha \simeq \Mov{-}0.88$ for NGC5949 and $\alpha \simeq \Mov{-}0.80$ for NGC6689. We illustrate this in Fig.~\ref{fig:RCobs1}: its top left panel confronts NGC5963's DM halo RC (black dots with error-bars) to RCs computed from \begin{inparaenum}[\itshape a\upshape)]
    \item the \citep[][]{DelPopolo2009} model that accounts for baryonic physics (yellow dashed line)\label{enu:dp09model},
    \item a fitted NFW profile (cyan line),
    \item a fitted pseudo-isothermal profile (ISO, short-dashed magenta line),
\end{inparaenum}
while its top right panel confronts the same kinds of fits to the RC from NGC5949, and the bottom panel only displays NGC2976's RC with a flat power law (black line) and model {\it\ref{enu:dp09model})} (dashed line). The NGC5963 RC is well fitted by the cuspy NFW profile's. The NGC5949 RC is equally well 
\noindent approximated by RCs from a cuspy NFW or 
ISO cored profiles. Finally, the NGC2976 RC reflects a very flat inner profile ($\alpha \simeq 0.01$). The three RCs displayed in Fig.~\ref{fig:RCobs1} all agree with the \cite{DelPopolo2012a} model.

It appears then that if a cored profile agrees with RCs of many dwarfs, some remain clearly cuspy.

In addition, opposite conclusions on the core/cusp question have been drawn from the same objects with similar techniques by different studies: \cite{Simon2003} analysed NGC2976's DM inner profile to lie within $-0.17<\alpha<-0.01$, when the same object yielded sometimes incompatible profiles in \citep{Adams2014}, traced either by gas, $\alpha=-0.30 \pm 0.18$, or stars, $\alpha=-0.53 \pm 0.14$.

In agreement with the preceding considerations, galaxies' observed RCs were recognised by \cite{Oman2015} to exhibit much more diversity than what simulations produced.

The conclusion this brings emphasizes the difficulty of galaxies' inner slope assessment, including for dwarfs. This moreover calls for a redefinition of the CC problem in terms of inner mass, rather than inner slope of galaxies density or RCs. Indeed, as shown by several studies, including those mentioned here, and despite kinematic maps recent improvements, there exists no agreement on DM slope distributions correlated with morphologies
\citep{Simon2005,oh,Adams2014}, but rather a range of profiles.

This lack of agreement appears even more obvious for the extreme ends of the mass spectrum of galaxies: for larger masses, dominated by stars, as in spiral galaxies, and for lighter galaxies, such as dwarf spheroidals (dSphs), which models require the account of biases, that result in opposite consequences.

To interpret and assess these results on dSphs, a range of approaches have been tried.
As stellar orbit anisotropy and mass are degenerate in the spherical Jeans equation \citep{Evans2009}, its results highly vary with assumptions. Similar degeneracies also hinder the approach applying parameter space maximum likelihood to Jeans modelling \citep{Wolf2012,Hayashi2012,Richardson2013}.
With Schwarzschild modelling, cored profiles are found~\citep{Jardel2012,Breddels2013,Jardel2013b,Jardel2013a}, for, e.g., Sculptor or Fornax dSphs. More recently a study by \cite{Hayashi2020} has found that the majority of dSPhs favor cuspy profiles. 
Multiple stellar populations methods also obtained cored profile
\citep{Battaglia2008,Walker2011,Agnello2012,Amorisco2012} for Fornax (measuring its slope at $\simeq$ 1 kpc) and Sculptor (for its slope at $\simeq$ 500 pc). Nevertheless, the lack of agreement on a single inner distribution for dSphs still remains, with contradictory results such as the cusp obtained from  Schwarzschild modelling for Draco \citep{Jardel2013a}.

Such scatter preventing a unified understanding occurs as well for galaxy clusters scales. In clusters MACS1206, MS2137-23, RXJ1133, A383, A1201, A963 and RXJ1133, studies with a combination of BCG's (Brightest Central Galaxy) stars velocity dispersion, weak and strong lensing, Sand et al. \protect\cite{sand1} found cores, and only the RXJ1133 profile could be compatible with the NFW model. Such flatter profiles were also obtained, in similar studies (see also \citep[][]{dpclust}), for A611 by Newman et al.~\protect\cite[2009]{newman}, for A383 by \mbox{Newman et al.}~\protect\cite[2011]{newman1} and for MS2137, A963, A383, A611, A2537, A2667, A2390 by Newman et al.~\protect\cite[2013]{newman2}. This contradicts other studies such as the cuspy profile in the strong lensing and X-ray observations of A611, among seven cuspy or flat profiles of relaxed massive clusters with average slope $\alpha=0.50 \pm 0.1$, by \mbox{Donnaruma et al.}~\protect\cite{donnaruma}.

More generally, conflicting estimates  with simulations are common with gravitational lensing \protect\cite{dahle,gavazzi2,donnaruma}, although much shallower slopes ($-$0.5) can also result from that technique \protect\cite{sand2,sand1,newman,newman1,newman2,DelPopolo:2002nd,DelPopolo:2005sq}. A similar wide slopes spanning can stem from X-ray analyses, ranging from cores at $-$0.6~\protect\cite{ettori}, \noindent to NFW-like $-$1.2 \protect\cite{Lewis2003}, up to very cuspy profiles $-$1.9 \protect\cite{arabadjis1}. However, NFW also agrees with some X-ray profiles \protect\cite{schmidt}.

Although the situation improved with more recent, higher resolution, observations, as they agree on flatter-than-NFW profiles on average, compared with the early data's conflicting density profiles 
inner structures, such high resolution pictures still display galaxy to galaxy inner structure diversity, reflected in \cite{Oman2015} simulations.

\begin{figure}
\begin{center}
\includegraphics[width=7cm]{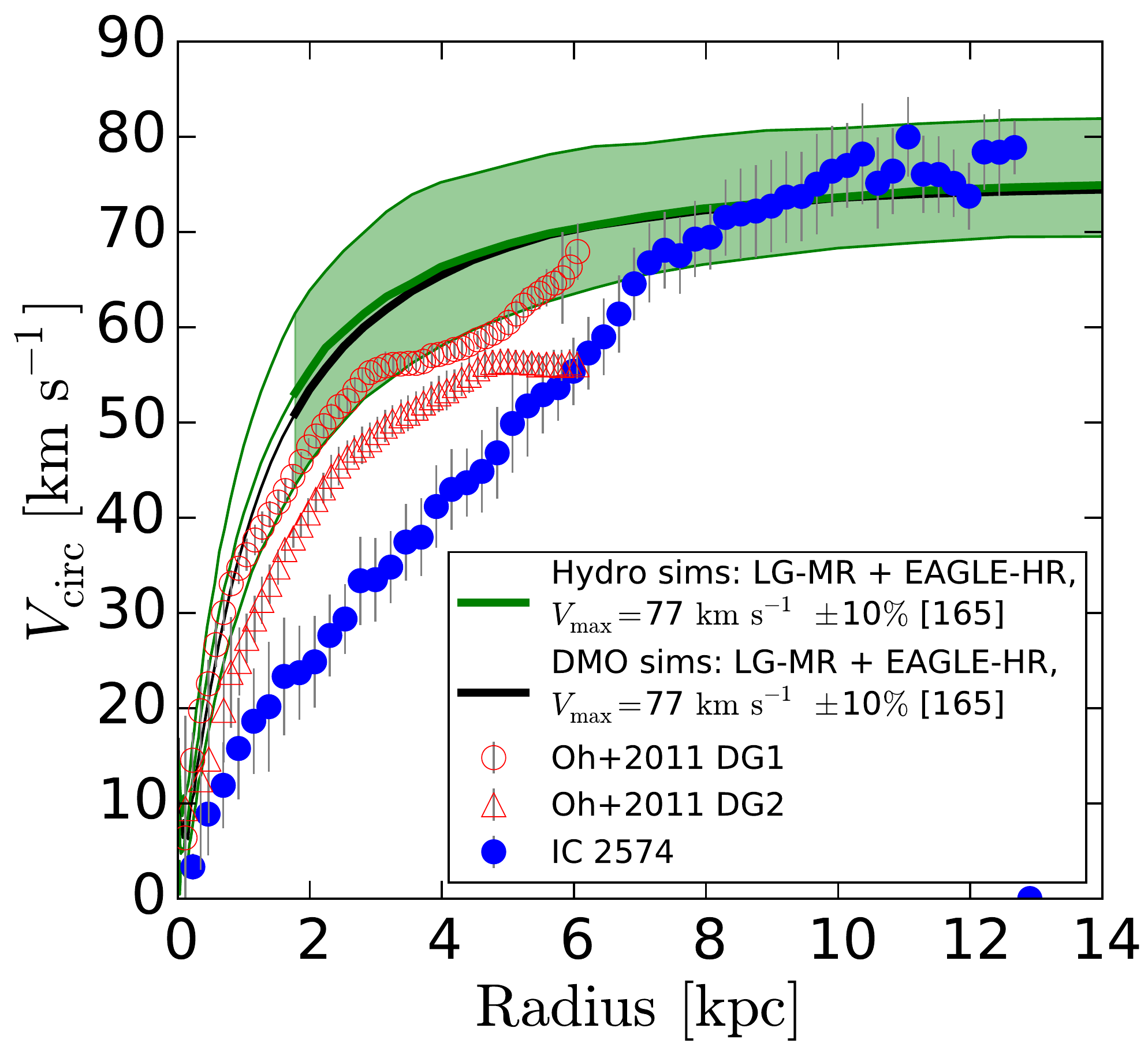}
\end{center}
\vspace{-12pt}
\caption{Observed RC from galaxy IC2574 (filled blue circles with error bars) compared with the DG1 (open red circles) and DG2 (open red triangles) simulated galaxies from \citep{Governato2010}. The green shaded region and line, as well as the black line, respectively stands for the scatter of one and both median RCs for simulated galaxies from Ref.~\citep[][which this figure, Fig.1, is extracted from]{Oman2015}.
}
\label{fig:RCsim1}
\end{figure}

Despite the agreement of the IC2574 galaxy's RC with the simulations from Ref.~\cite{Oman2015} (presented in Fig.~\ref{fig:RCsim1}, shaded green band) for radii $> 6$ kpc, the galaxy's inner region behaviour differs completely from their simulations. Such divergence emphasizes that correct RC slope predictions for some inner regime doesn't entail that the successful simulation remains accurate in all regimes of the RC. As the main characteristic of the CC problem resides in the inner galaxy's mass deficit \cite{Oman2015}, the focus should lie in the DM excess in the central distribution produced by CDM rather than on the density profile shape. The observation vs prediction discrepancy already appears where the asymptotic value of the circular velocity is reached \cite{kuzio_spekkens}.

It is therefore unfortunate that practically no observational papers in the field depart from the inner slope estimation method using $\alpha$. A notable exception can be found in the inner slope estimation via the integrated indicator $\Gamma \equiv \Delta\log{M}/\Delta\log{r}<3-\alpha$ performed by \cite{Walker2011} in the cases of the Fornax and Sculptor galaxy clusters. As $\Gamma$ is integrated, contrary to the local $\alpha$, its evaluation is easier, but provides a lesser constraint on the inner slopes.

\Mov{Finally, an alternate probe of the DM inner profile flattening to rotation curves consists in the dynamical friction timing (DFT) technique, that is assessing the time an object like a globular clusters (GC) requires to move inside the distribution of mass. It clearly represents an indirect method to obtain the DM inner slope, however it presents the advantage of directly probing the density, rather that the integrated mass furnished by rotation curves. Thanks to the presence of GCs in dwarf galaxies, like Sagittarius or Fornax, \citep{Hernandez1998} was able to show analytically that the inner density profile must be cored. This behaviour focused on GCs  was confirmed by several simulations \citep{Sanchez-Salcedo2006,Read2006,Inoue2009,Chowdhury2020}. The key interests of the DFT technique are twofold: \begin{inparaenum}[\itshape 1\upshape)]
\item it concentrates on smallest scale galactic systems like dwarf galaxies, on which the proposed solutions to the CC problem show their main drawbacks, and
\item its method is independent from that of rotation curves, giving information on the structure of the mass distribution in the studied systems.
\end{inparaenum}
Recently, DFT studies have shifted their focus from GCs to DF on stars. Using analytical approaches, as in Hernandez \cite{Hernandez2016}, or numerical techniques, in Inoue \citep{Inoue2017}, on ultra-faint satellite galaxies of our galaxy, was shown that DM fractions imply DFT for the stars themselves and are incompatible with cusped dark matter haloes, at these tiny scales, out of reach of SNF models, and even maybe of DFBC solutions.  
}

Since smaller mass objects are more likely to harbour inner profiles in agreement with predictions of dissipationless, N-body, codes (cuspy), a definite verdict on the nature of the inner slope, cored or cuspy, of the smaller end of the mass spectrum, i.e. dSphs, is critical.

\section{Initially proposed CC problem solutions}\label{sec:earlySol}

In the wealth of propositions to solve the CC problem, the initial attempts were determined by the limited state of knowledge and observations. Systematic effect and poor resolution were first invoked as the origin for the CC inconsistencies \protect\citep[e.g.][]{vandenbosch,vdbswater}: they blamed beam smearing, slit-misplacement, non-circular motions or beam off-centering, as they are sources of systematically lowering slopes.

For instance, finite HI emission beam size tends to smear the HI observations, inducing a larger measured disk, depending on the HI distribution, beam size, intrinsic velocity gradients and galaxy inclination angle. With modern high spatial resolution ($<$1 kpc, see below), the smear can be resolved.

The slit misplacement for an H$\alpha$ emitting galaxy can entail missing the galaxy's dynamic centre, inducing a flatter measured profile. One of the solutions to avoid this slope underestimate have used \mbox{3d spectroscopy~\citep[][]{kuzio08,Spano2008}.} As H$\alpha$ observations assume circular motions for the gas, any non-circular orbits underestimate the slope. Some have found \cite{trachte} that those departures from circular velocity remain within the order of a few km/s. Current high resolution rotation curve data are able to discriminate cuspy from cored haloes with asymptotic inner slopes~\protect\cite{KuziodeNaray2011}.

Projection effects were put forward by several papers \protect\cite{power,navarro04,hayashi} to explain the discrepancy of their simulations with observations, despite their claims of good agreement of their profiles, becoming shallower in the range the virial radius towards the centre. Their argument lies on the practice of representing the 3D DM haloes motion with a spherically averaged circular velocity, traced by the gas disk rotation speeds. Their claim is to blame the disagreement with data on the correspondence fitting formulae rather than their simulations  \protect\cite{hayashi}. Current DM-only high resolution simulations clearly reject this explanation, as their inner profiles cannot reach shallower slope than  $ \simeq -0.8$ \protect\cite{Stadel2009}, compared with high resolution galaxy observations revealing much smaller values. 

More generally, simulations limitations, such as over-merging or lack of relaxation or resolution, have been blamed to source the CDM model's failure  \protect\cite{deBlok2003,deblok3,borriello}. This was already questioned then by Ref.~\protect\cite{diemand}'s convergence tests proving the dissipationless N-body simulations of the time were adequate to properly obtain the CDM density profiles, despite their neglecting the galaxies inner kpc regions effects of baryons. Such effects were shown to dominate the DM generated profiles in clusters' central 10 kpc regions \protect\cite{sand1,newman,newman1,newman2}. Modern simulations no longer suffer from the blamed limitations and hydrodynamic cosmological high resolution models now account for baryons, as will be discussed in the next section on baryonic resolutions of the CC Problem. 

Questioning of the very paradigm of CDM opened the door to speculations on the nature of DM, such as warm protect\cite[WDM][]{som_dol}, fluid \protect\cite{peebles2000}, repulsive \protect\cite{goodman}, fuzzy \protect\cite{hu}, decaying~\protect\cite{cen1}, self-annihilating \protect\cite{kap}, or self-interacting \protect\cite{sperg_ste} DM. More fundamental questioning led to altered small scale DM power spectrum \citep[e.g.,][]{Zentner2003} or modifications of gravity itself, such as the so-called $f(R)$ \citep{Buchdahl1970,Starobinsky1980} or $f(T)$ \citep{Bengochea2009,Linder2010,Dent2011,Zheng2011} theories, or the so-called Modified Newton Dynamics \citep[MOND,][]{Milgrom1983b,Milgrom1983a}\footnote{
Generalisations of Einstein's General Relativity, $f(R)$, and $f(T)$ theories modify gravity in distinct ways. Replacing, in their Lagrangian, the Ricci scalar by a function $f(R)$, the first type dates from the Buchdahl~\protect\cite{Buchdahl1970} introduction in 1970 but picked up interest with Starobinsky~\protect\cite{Starobinsky1980} a decade later. Developed from  the Teleparallel Equivalent of GR, $f(T)$ theories also modify their Lagrangian from the torsion scalar to a function. On one hand, both types were introduced to absorb the origin of cosmic acceleration into gravity and dispense from the need for DE \protect\citep[][]{Bengochea2009}. On  the other hand, MOND, introduced to fit galaxies rotation curves by Milgrom in 1983~\protect\cite{Milgrom1983b,Milgrom1983a}, purports to replace DM.
\Mov{The DM effects of the MOND Newtonian modification also lead to other GR generalisations with the same aim \cite{Mendoza:2012dh,Campigotto:2016cmt,Barrientos:2018cnx,Barrientos:2018giw}}}. Such paradigmatic modifications, whether touching DM nature, initial conditions or the nature of gravity, will hereafter be denoted, in their impact the CC problem, as "cosmological solutions", and developed in Sec.~\ref{sec:CosmoSol}.

\section{Baryonic Resolutions of the CC Problem}\label{sub:BsolCC}

Before resorting to the above mentioned ``cosmological solutions'' to deal with the CC problem, a thorough check, whether neglected or poorly understood local physics could explain it, would be wise. Indeed, discarding the concordance, i.e. $\Lambda$CDM, model would risk to significantly alter the otherwise very successful predictions $\Lambda$CDM provided, thus loosing the explanations for many observations of our Universe. This is the aim of the current section.

Keeping the framework of the $\Lambda$CDM model, ``astrophysical solutions'' alter the results of pure CDM structure formation with some ``heating'' mechanism applied to inner DM distribution, originated in the effects of baryons. The inner density profile gets then flatter than for pure CDM evolution. Such ``heating'' mechanism could be provided by
\begin{enumerate}
\item accounting for the action, in structure formation, of angular momentum;
\item the induced dynamics from a central black hole, claimed to shallow the DM cusp~\protect\cite{delliou_b,delliou_b1,delliou_b2}; 
\item the modification of equilibrium from the presence of a rotating bar;
\item the mechanism of dynamical friction that induces baryons-DM transfer of angular \mbox{momentum  \protect\cite{ElZant2001,ElZant2004,DelPopolo2009};}
\item the supernova (SN) explosions-generated bulk motions in gas, dubbed AGN  feedback~\protect\cite{Mashchenko2006,Mashchenko2008,Governato2010}.\label{enu:SNfeedback}
\end{enumerate}

Larger angular momentum acquired by a proto-structure has been correlated to flatter inner density profile by many contributors within the framework of spherical~\protect\cite{nusser,hiotelis,delliou,ascasibar,williams,dpl,dpl1,dpl2,dpl3,dpl4,dpl5,DelPopolo:2008mp}, or even elliptic~\protect\cite{Popolo:2002mp}, infall models.\footnote{A General Relativistic version of secondary infall models is at the root of works by a group including Mimoso and Le~Delliou~\protect\cite{mimoso,mimoso1,delliou_she,delliou_she1}.} Such models were able to produce RCs agreeing with those of dwarf galaxies~\protect\cite{williams}, which profile builds flatter than giant galaxies' because structure's density peak height\footnote{A proto-structure's peak height measures the ratio of its central peak overdensity $\delta(0)$ to its mass variance  $\sigma$ \citep[see][]{DelPopolo1996}, as $\nu=\delta(0)/\sigma$. More  massive objects are characterised by larger $\nu$.} anticorrelate with the angular momentum they get during their formation. 

Dynamical friction was shown to flatten, and even erase, the DM formed cusps in both galaxy clusters and dwarfs by transferring energy from baryon clumps to DM~\protect\cite{ElZant2001,ElZant2004}. Conversely, DM adiabatic contraction (AC) in haloes steepens their density profiles~\protect\cite{blumenthal,gnedin}.\footnote{As computed with iterative methods \citep[e.g.,][]{spedi}.}

The above effects \citep[see also][]{DPK2009} were all combined in the model from Del Popolo~\protect\cite{DelPopolo2009}, including:
\begin{itemize}[-]
\item tidal torques caused ordered angular momentum acquired during the structure formation;
\item its obtained random angular momentum;
\item its dynamical friction induced baryons-DM energy and angular momentum exchanges;
\item and the effects of its DM adiabatic contraction (hereafter, AC).
\end{itemize}
\begin{figure}
\centering
\includegraphics[width=8cm]{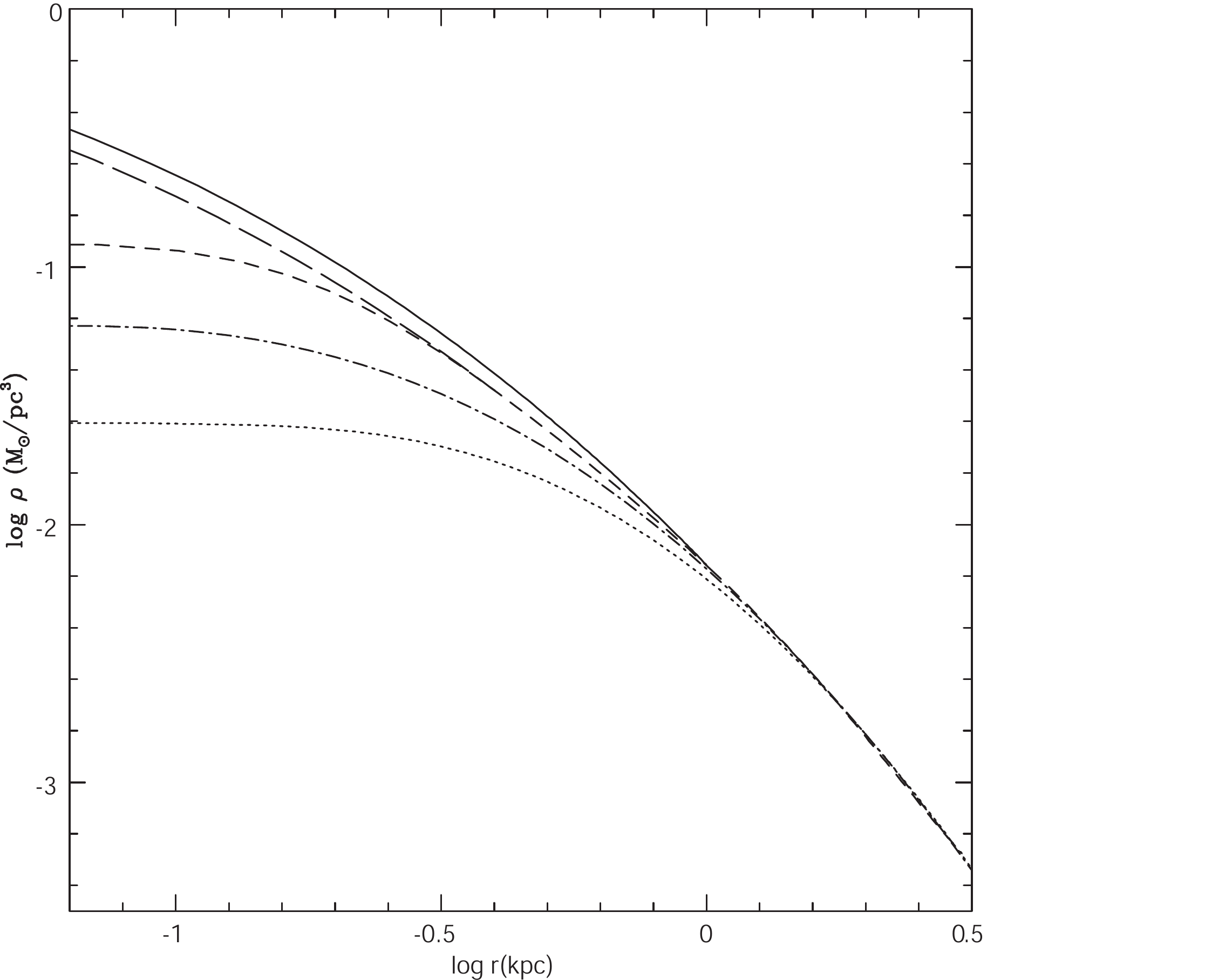}
\includegraphics[width=6.8cm]{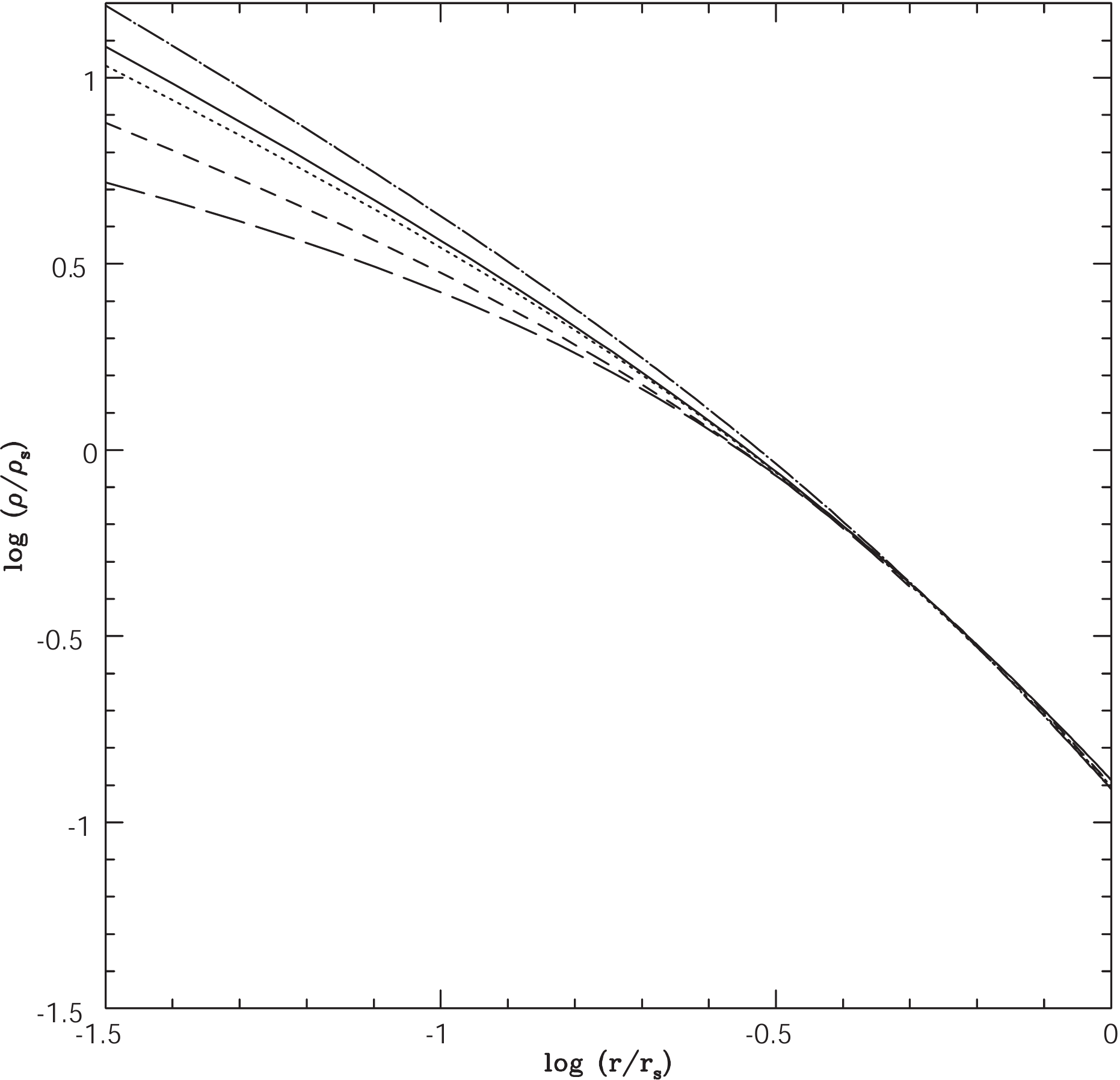}
\caption{Halo density profile evolutions with redshift \cite[following the DFBC scenario of][]{DelPopolo2009}. The evolution for a $10^9 M_{\odot}$ halo yields profiles at $z=10, 3, 2, 1$, and 0, characterised resp. by the solid, long-dashed, short-dashed, dot-dashed and dotted lines, and presented in the left panel. The profile evolution, ending at $z=0$ with the total (DM+Baryons) mass density profile of a $10^{14} M_{\odot}$ halo and marked by the dot-dashed line, is reproduced in the right panel. That DM profile is characterised at $z = 3, 1.5,$ $1$~and  0 resp. by the solid, dotted, short-dashed, and long-dashed lines \citep[this figure is extracted from][Figs.~3 and 5]{DelPopolo2009}.
}
\label{fig:DFevolDen}
\end{figure}
In addition to inducing flatter density profiles in the center of galaxies and clusters, dynamical friction and angular momentum also tend to modify their overall internal structure \cite{dpclusta,dpclustb,dpclustc,DelPopolo:2003iu}.
The simple confrontation of real structures, including baryons, with the outputs of dissipationless simulations is demonstrably incorrect ~\protect\cite{DelPopolo2009}: the non-negligible action of baryons in the inner regions of structures explains the failures of N-body simulations to reproduce observed density profiles. This is illustrated in Fig.~\ref{fig:DFevolDen}, showing halo evolution for two masses, $10^9$ and $10^{14} M_{\odot}$, modeled using the ``Dynamical Friction from Baryonic Clumps'' scenario (hereafter DFBC scenario) from Ref.~\protect\cite{DelPopolo2009} and discussed below.

The impact of baryons content, environment and formation history on dwarf galaxies density profiles was demonstrated with the Del Popolo DFBC model \protect\cite{DelPopolo2012a}.

The ``heating'' mechanism \ref{enu:SNfeedback} was first put forward by Navarro et al. \cite{Navarro1996a}, flattening the inner profile from the feedback of SN explosion into the medium, as discussed below.

In the present state of the field, the discussion on astrophysical solutions to the CC problem have narrowed down to the following mechanisms 
\begin{enumerate}
\item the cusp reduction through ``supernovae feedback flattening'' (SNFF)
\citep{Navarro1996a,Gelato1999,Read2005,Mashchenko2006,Mashchenko2008,Governato2010,Governato2012}, and
\item the cusp flattening via the ``Dynamical Friction from Baryonic Clumps'' (DFBC) 
\citep[][]{ElZant2001,ElZant2004,Ma2004,Nipoti2004,RomanoDiaz2008,RomanoDiaz2009,DelPopolo2009,Cole2011,Inoue2011,Nipoti2015}.
\end{enumerate}

They will be discussed in more details in Secs.~\ref{sub:SNsolCC} and \ref{sub:DFclumps}, and are put in context below.

In general, baryons and DM usually ``interact'' in smooth distributions via AC, that can be counterbalanced by transferring energy back from baryons to DM. AC leads DM to collapse towards the centre of the halo, steepens its profile and increase its central density. The counteracting transfer has the opposite effect. It can occur via
\begin{enumerate}
\item Dynamical friction transfer of incoming clumps orbital energy to DM. \\This leads DM particles to drift
towards the halo edges and thus to flatten the DM density profile
\citep[see also the review of \cite{Pontzen2014}]{ElZant2001,ElZant2004,Ma2004,Nipoti2004,RomanoDiaz2008,RomanoDiaz2009,DelPopolo2009,Cole2011,Inoue2011,Nipoti2015}.
\item DM particles ``heating'' from galactic {\it internal} energy sources \citep{Pontzen2014}.\\ Such source expelling from, or relocating baryons in, the halo (for instance, bulk motions induced from SN explosions,~\citep[][]{Mashchenko2006,Mashchenko2008}) can temporarily flatten the gravitational potential, shift DM particles outwards, and flatten the cusp.
\end{enumerate}

We will now discuss the two favoured ``heating'' mechanisms in the CC problem solution, and how those solutions are synthesised into the proposal of a mass dependent density profile for DM haloes with baryons.

\subsection{Discussion on Supernovae Feedback Flattening}\label{sub:SNsolCC}
Historically, the first solution proposed to the CC problem while accounting for the effects of baryons, as initially suggested in Ref.~\cite{flores} and confirmed by many efforts, employed SNF \citep{Navarro1996a,Gelato1999,Read2005,Mashchenko2006,Mashchenko2008,Governato2010,Governato2012,Teyssier2013,Sawala:2015cdf}.

Profile flattening by a single sudden baryons expulsion into the halo was first shown by Ref.~\cite{Navarro1996a}, and found to be most effective in dwarfs and other shallow potential structures. However, energy from a single explosion was shown \cite{Gnedin2002} to remain insufficient to drive the profile to a core, while repeated moderately violent explosive events would be required \citep[see][for a different point of view]{GarrisonKimmel2013}.

A more detailed study of the \cite{Navarro1996a} approach, focused on evolving from NFW profiles to DDO154's RC via a gas outflow event modeled from an abrupt change in the disk potential, by Gelato and Sommer-Larsen \cite{Gelato1999}, established that at least 75\% of the disk mass should be expelled that way to generate this RC.

In the case of larger galaxies and deeper potential structures, a core was shown by Read and Gilmore \cite{Read2005} to result from repeated and alternate outflows and re-accretions of gas.

In primordial galaxies, a core was determined to follow from SN-explosions-driven gas random bulk motions \cite{Mashchenko2006,Mashchenko2008}, as confirmed in \cite{Governato2010} simulations. The average slope of the latter simulations were compared with observations of THINGS dwarfs in Refs.~\cite{Oh2011a,oh}. Producing simulations for larger galaxies than \cite{Governato2010}, and comparing them with observations \cite{Governato2012}, correlation between the stellar mass $M_{*}$ and the inner slope were obtained for~$M_*> 10^7 M_{\odot}$ galaxies were obtained. These results were obtained from simulated galaxies thanks to the N-Body+SPH code GASOLINE \citep{Wadsley2004}. With an 86 kpc softening, gas particles and DM particles resolution were reduced to $M_{\rm p,gas} = 3 \times 10^3 M_{\odot}$ and $M_{\rm p,DM} = 1.6 \times 10^4 M_{\odot}$, reps., thanks to the ``zoom'' technique of Ref.~\citep{Katz1993}. Two realisations were produced, distinguished by their star formation thresholds: a High Threshold run (HT), with star forming hydrogen cloud densities $>$$100/$cm$^3$, and a Low Threshold run (LT), with hydrogen densities $>$$0.1/$cm$^3$. Both SNF blast wave mechanism \citep{Stinson2006} and/or early stellar feedback \citep{Stinson2013} were implemented in those simulations, as later by \cite{DiCintio2014} in the same way. The Governato simulations resulting energy feedback to the interstellar medium (ISM) amounted to $10^{51}$ ergs from $>$$8$$M_{\odot}$ stars, while the ISM gas ejecting energy was coupled to the SN expelled energy via the coefficient $\epsilon_{\rm esf}$. The fiducial $\epsilon_{\rm esf}=0.1$ was selected for the MaGICC simulations \citep{Stinson2013}.

\begin{figure}
\hspace{2cm}
\includegraphics[width=12cm]{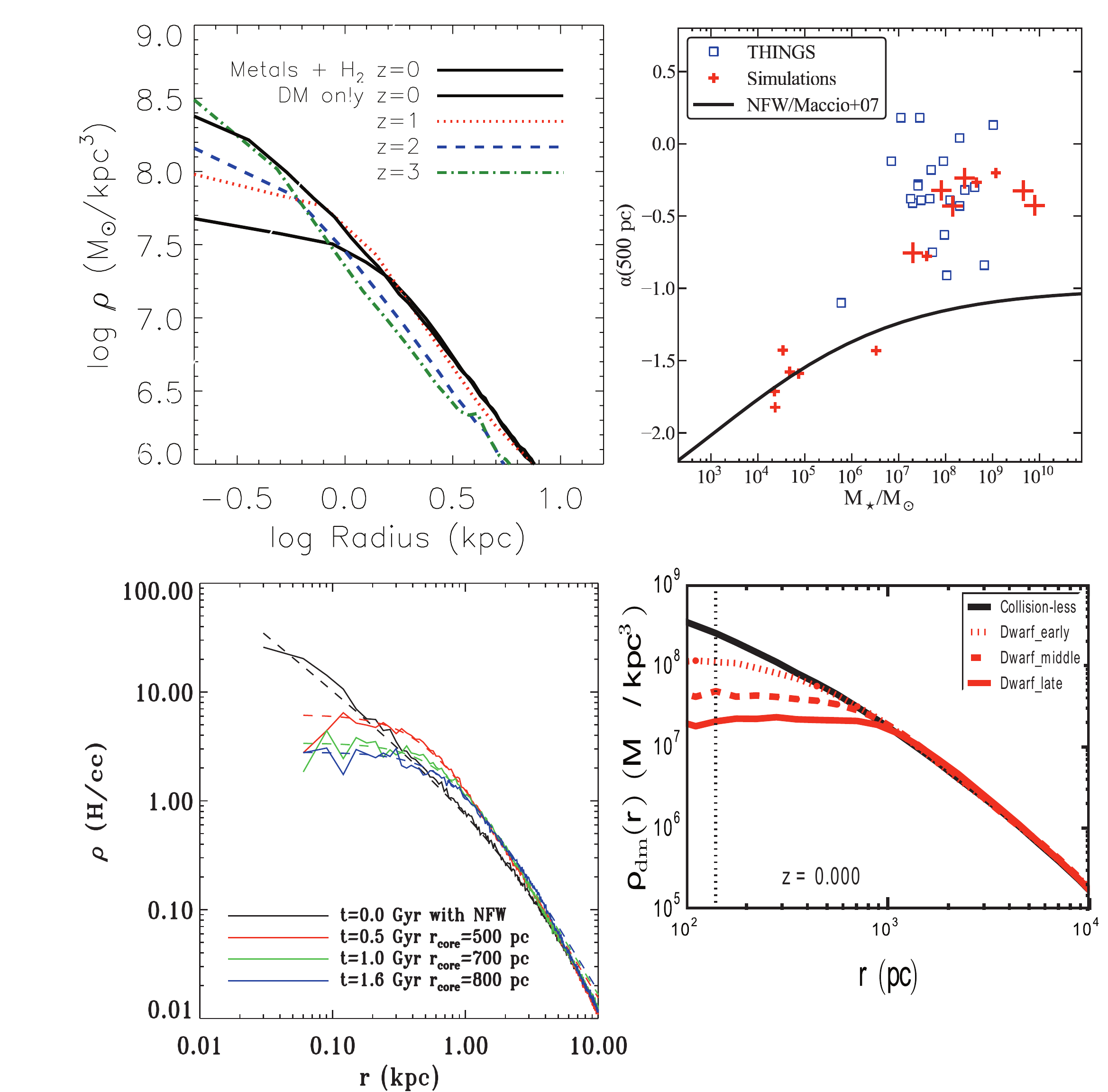}
\caption{Density profiles including baryonic effect. The Ref.~\cite[Fig.~1]{Governato2012}'s hydrodynamic simulations present density profile evolution (top left panel) reflected on their inner logarithmic slopes (at 0.5 kpc). The Ref.~\cite{Governato2012}'s results (red crosses) are compared, in the top right panel, for different stellar mass galaxies, with THINGS galaxies' results (open squares), and with DM-only N-body simulation \citep[reproduced from][Fig.~3]{Governato2012}, shown as a solid line. Results from hydrodynamic simulations are presented in the bottom panels: density profile evolution from Ref.~\cite[Fig.~5]{Teyssier2013} are shown on the bottom left panel, while the bottom right presents hydrodynamic evolution for three different types of dwarfs, where stars all form either in early times, in medium or in late times \cite[Fig.~6]{Onorbe2015}.
}
\label{fig:DenHydro}
\end{figure}

Comparably, a combination of bursty star formation and SNF, inducing inner (1 kpc) halo potential fast oscillations and expanding gas bubbles when the central cold gas density reaches $>$100/cm$^3$ and stars start to form\footnote{This process agrees with the assumptions from Ref.~\cite{Governato2010}; the bulk gas flows start beyond $>$10/cm$^3$ \citep{Ceverino2009}.}, was shown to flatten the cusp \cite{Pontzen2012}. However in this model, no noticeable modifications of the DM inner density profiles are induced for smaller densities (e.g., 0.1/cm$^3$). The case of larger mass galaxies was treated by Governato et al.~\cite{Governato2012} with the same method (also see Fig.~\ref{fig:DenHydro}, top panels). Agreement with the results of \citep[][]{Pontzen2012} (Fig.~\ref{fig:DenHydro}, bottom left panel) was obtained from introducing a new SNF scheme in the RAMSES adaptive mesh refinement code by Teyssier et al.~\cite{Teyssier2013}, showing flat galaxies inner profile for $M_*> 10^7 M_{\odot}$. Similar outcomes with a lower threshold at $M_*> 10^6 M_{\odot}$ were produced by Onorbe et al.~\cite{Onorbe2015} (Fig.~\ref{fig:DenHydro} bottom right~panel)\footnote{\label{ftn:coreMassLimit} Essentially, $M_*< 10^6 M_{\odot}$ galaxies are not able to turn cusps into cores under the SNF mechanism.}.

Since their simulations found little effect of baryons on rotation curves of galaxies with $V_{\rm max}<60$ km/s, even in their inner regions, this perspective was denounced by \cite{Oman2015}. They argued that the cores obtained by \cite{Pontzen2012}, since no evidence of core formation was detected in their \cite{Oman2015} work, were inherently reflecting an {\it ad hoc} choice of parameters.

Such understanding agrees with the findings of Ref.~\cite{GonzalezSamaniego2014}, which 7 high-resolution dwarfs simulations, included in halos with $1-2 \times 10^{10} M_{\odot}$ and following different assembly history, could not reveal any inner core flattening. A very extended assembly history dwarf, thus also having an extended star formation rate (SFR) history, exhibits their lowest detected inner slope: they measured it to reach, at 0.01--0.02~$R_{\rm vir}$, up to $-$0.8. It also remains consistent with the core-less simulated realistic galaxies of Ref.~\cite{Schaller2015}.

In addition to those contradictions in core formation between various high resolution hydrodynamical simulations, several criticisms have arisen on the methods and assumptions used in simulation models of the types of \citep[][]{Governato2010,Pontzen2011}, as they obtain faulty
\begin{enumerate}
\item core formation energetics: the number of stars, obtained in their $M_*<10^7 M_{\odot}$ galaxies, generates an insufficient energy to flatten the galaxies cusp \citep[][]{Penarrubia2012}. Moreover, their required core-forming baryonic mass marginally exceeds the dSphs observed baryon content \citep{GarrisonKimmel2013}. That problem in Penarrubia \cite{Penarrubia2012} is illustrated in Fig.~\ref{fig:CCenergy}'s left panel. However, the right panel displays the opposite results from Maxwell's study.\\
In addition, the SNFF solution they propose for the CC problem, requiring the formation of numerous SNs, entails a star formation efficiency (SFE) constraint towards larger values, opposite to the SFE constraint induced by the complementary solution to the Too-Big-To-Fail (TBTF) problem \cite{boyl,boyl1}\footnote{``Too Big to Fail'' is used in the context of Milky Way (MW) satellite simulations producing bigger objects than observed MW satellites without any mechanism explaining why such object would fail to be detected.} 
\begin{figure}
\begin{center}
\includegraphics[width=6cm]{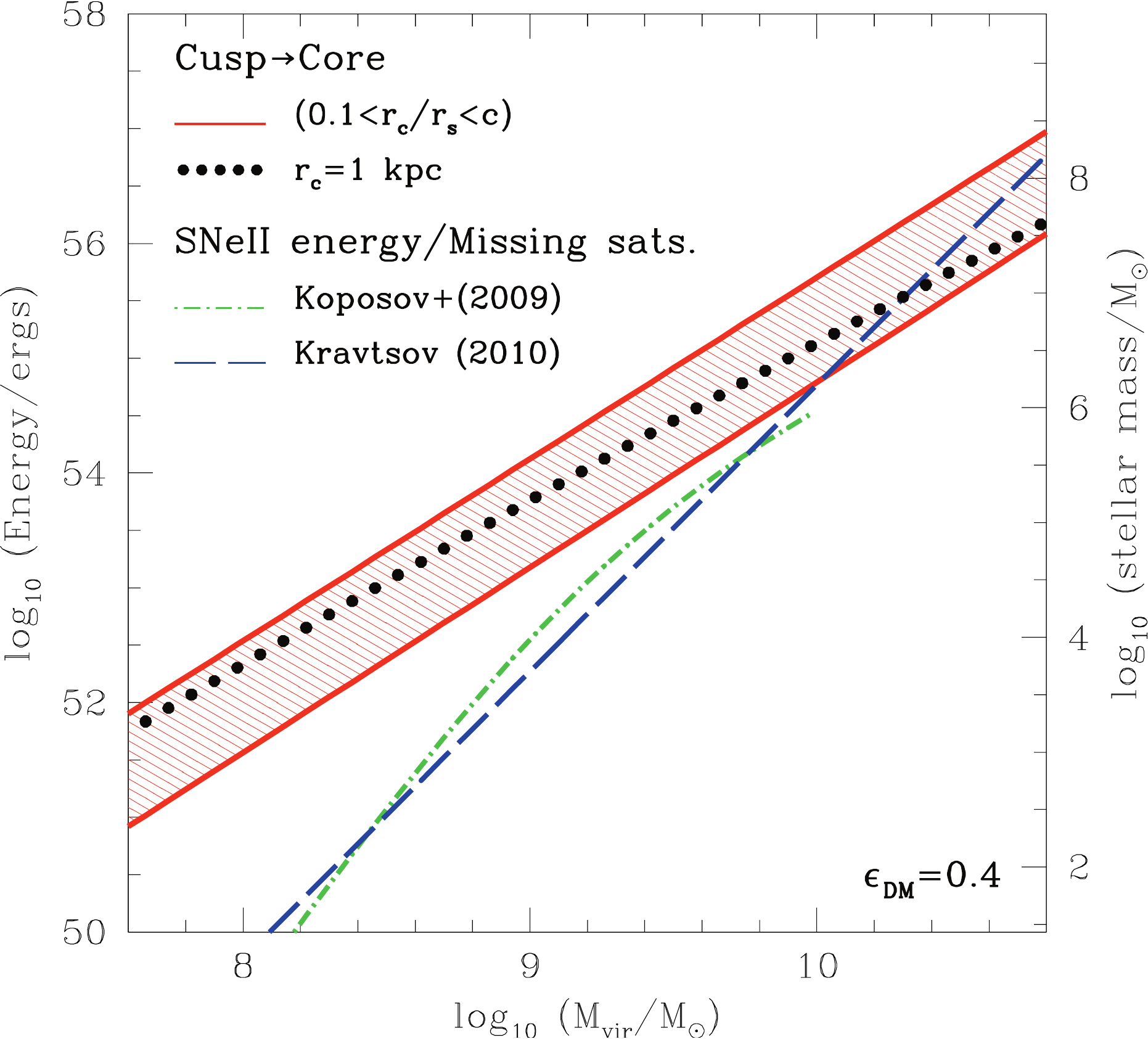}
\includegraphics[width=6cm]{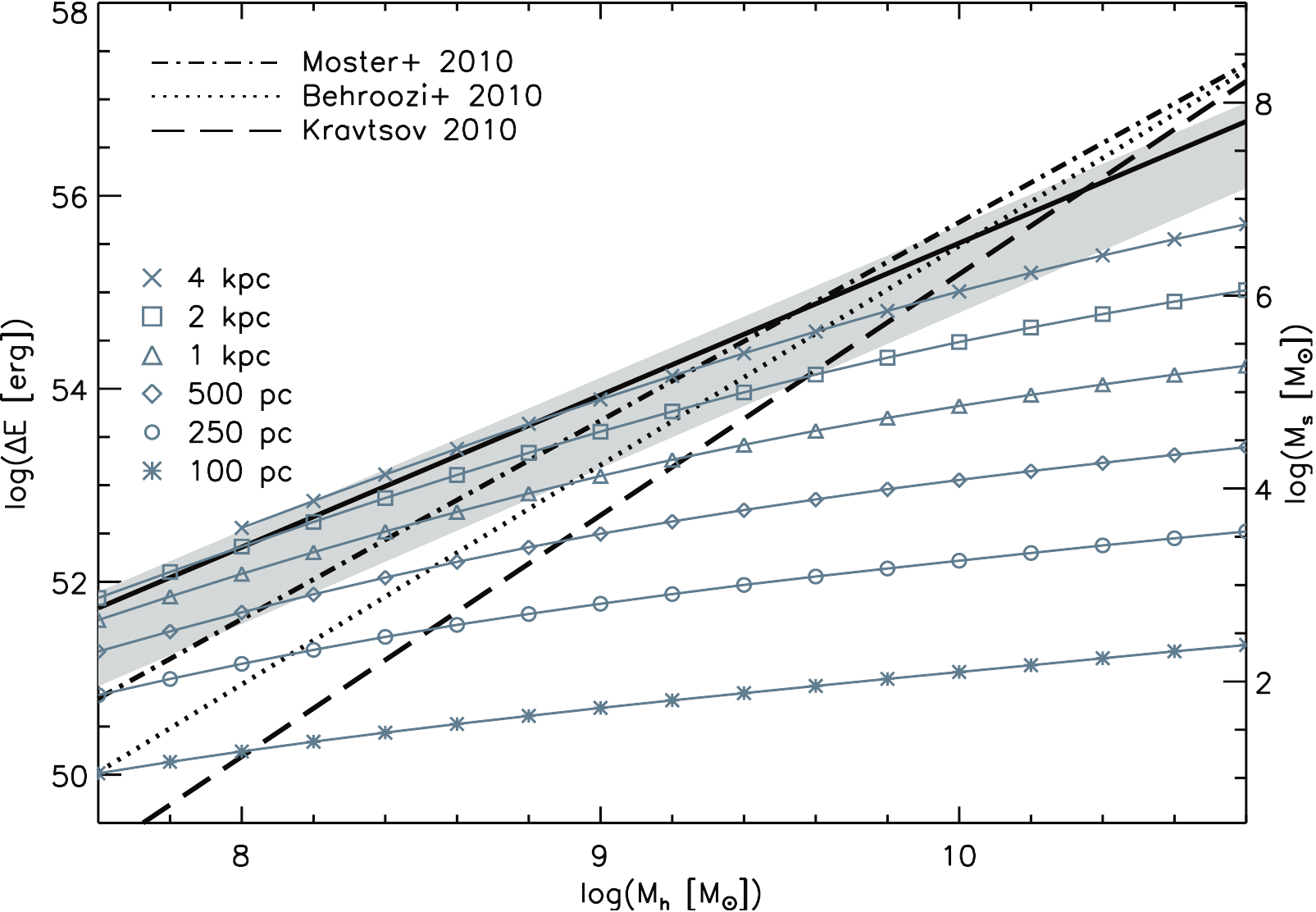}
\end{center}
\vspace{-6pt}
\caption{ 
Minimal SN energy/mass output, $\Delta E$, needed to convert a DM cusp into a core as a function of halo mass, $M_h$, measured at virial mass, $M_{\rm vir}$, in the left panel. The constraints on the minimal amount of energy as a function of mass can also be expressed in terms of stellar mass, standing for luminosity, shown in the right axes of each panel. Conversion from stellar mass proceeds, in the left panel, from a luminous MW satellites constrained star formation efficiency $F_* = F_*(Mvir)$, the resulting luminosities converting into SNeII energy output via Ref.~\cite{Penarrubia2012}'s Eq.~(6) with strong energy coupling $\epsilon_{DM} = 0.4$, while it assumes, in the right panel, 100\% efficiency ($\epsilon_{DM} = 1$).\\
The left panel, reproduced from~\citep[][Fig.~2]{Penarrubia2012}, presents: 
{\it a)}~the range of minimum SN explosions energy required to obtain a core of size $r_c$ given within $0.1 <r_c/r_s < c$, with $c$ and $r_s$, respectively being the concentration parameter and the scale radius of the NFW profile at given virial mass $M_{\rm vir}$, as the red shaded area; 
{\it b)}~for a fixed core size $r_c=1$ kpc, the dotted black line is obtained; 
{\it c)}~the SN explosion energy outputs, indicated from two different studies~\cite{koposov,Kravtsov2010} as dot-dashed green and dashed blue lines, denote those compatible with the ``missing satellite'' \cite[][]{Klypin:1999uc,moore1} problem and reveal, for haloes with $M_{\rm vir} < 10 M_\odot$, the tension
with the ``core/cusp'' problem.\\
The right panel, extracted from \citep[][Fig.~3]{maxwell}, shows: 
{\it a)}~the left panel red shaded area of Penarrubia~et~al.~\cite{Penarrubia2012}'s energy estimates, as shaded grey area; 
{\it b)}~solid lines with symbols for the conversion energy of cuspy to pseudo-isothermal density profiles at fixed core sizes, indicated in the legend by the symbol types; 
{\it c)}~the solid black line for the $\Delta E$ scaling with halo mass $M_h$, while the ratio of the cusp mass redistribution limit radius, $r_m$, over the halo radius, $r_h$, is fixed;
{\it d)}~the dotted, dashed, and dot-dashed lines, respectively, for the Behroozi et al. \cite{Behroozi2010}, Kravtsov
\cite{Kravtsov2010}, and Moster et al. \cite{Moster2010} $M_\star-M_h$ relations.
}
\label{fig:CCenergy}
\end{figure}
\item energy coupling: their $\epsilon_{\rm SN} \simeq 0.4$ exceeds the 0.05 coupling that studies such as \cite{Revaz2012} deduced;
\item star formation threshold: the \cite{Governato2010} results have been found in \citep{Sawala2014b,Oman2015} to require a very high star formation threshold;
\item tensions with TBTF solution: those models simply encounter adverse conditions \citep{Ferrero2012,GarrisonKimmel2013,Papastergis2015} when trying to solve the TBTF problem;
\item feedback resolution: accurately describing the cusp transformation into a core via their feedback processes would require better resolution than they present \citep{Choi2014,Laporte2015a,Laporte2015b,Marinacci2014}.
\end{enumerate}

Despite the SNF limitations,\footnote{Recall footnote \ref{ftn:coreMassLimit}.}
for $M_{\ast}>10^6 M_{\odot}$, correlations between $M_{\ast}$
and galaxies inner slope have been found, comparing THINGS dwarves average slopes with \cite{Governato2010}'s simulations in Refs.~\cite{Oh2011a,oh}, and similarly in \cite{Governato2012} for larger objects.

Inversely, hydrodynamic simulations for $M_*<10^{6} M_{\odot}$ produce cuspy profiles, in contradiction with the Fornax and Sculptor cored inner structure found by Ref.~\cite{Walker2011} thanks to their more reliable DM haloes' slopes from mass profiles $\Gamma \equiv \frac{d \log M}{d \log r}<3-\alpha$. Only with 50\% more SN energy injections could  Ref.~\cite{Madau2014} claim agreement with \cite{Walker2011}'s cores.

Assuming SN feedback and other specific assumptions, a Semi-Analytic model from Gnedin and Zhao \cite{Gnedin2002} claimed impossibility to obtain cores. As, among details of stellar feedback, the core formation impact of baryonic clumps, such as in Ref.~\cite{DelPopolo2009}, is ignored, such objection is weakened.

However, there exists one core-forming SNF dwarf galaxies simulation for masses $<10^6 M_{\odot}$: with the P-SPH mode in the GIZMO code, \cite{Onorbe2015} obtained cores in a much less natural way than in the DFBC scenario, discussed in the next section.

\subsection{Discussion of Gas Clumps Dynamical Friction\label{sub:DFclumps}}

As discussed in the introduction of the present Sec.~\ref{sub:BsolCC}, Refs.~\cite{ElZant2001,ElZant2004} proposed an alternate path from SNF for the turning of cusps into cores: baryon clumps with mass ratios to the system above $\simeq\!0.01\%$ can transfer, through dynamical friction (DF), clumps orbital energy to DM. Such heating induces similar effects to SNF without the need to wait for a full massive stellar cycle, moving to outer orbits the central DM particles and thus flattening the inner DM density profile. It is more efficient on earliest, smallest haloes of the hierarchical structure formation.

DFBC was first shown to function in galaxies \cite{ElZant2001}, then in clusters \cite{ElZant2004}. Cluster C0337-2522 evolution with such model \cite{Nipoti2004} showed the BCG generation precedes DFBC inner slope formation, resulting in flatter than NFW inner profiles with $0.49<\alpha<0.90$.

Implemented in a hybrid N-body/SPH simulation, DFBC galaxy evolution showed heated up cusps by baryons subhalos into a core of $\simeq$1 kpc, and was compared with DM-only systems and mixed DM/baryons systems \citep{RomanoDiaz2008}.

Refs.~\citep{Cole2011,Inoue2011,Nipoti2015}'s simulations confirmed such outcomes.

The main steps of the DFBC can be outlined in the following: 
\begin{itemize}
    \item the linear phase develops from initial DM and diffuse gas proto-structures.
    \item DM gravitational collapse first form non-linear potential wells, attracting then the baryons. 
    \item Unstable accreting gas then fragment into clumpy structures \citep[e.g.,][]{Noguchi1998,Noguchi1999,Immeli2004a,Immeli2004b,Bournaud2007,Agertz2009,Aumer2010,Ceverino2010,
Ceverino2012}, leading to highly gas-rich discs.
    \item Further instability in the rotating discs, triggered by their surface density, $\Sigma$, exceeding a threshold at $Q \simeq \sigma \Omega/(\pi G \Sigma)<1$ \citep[with $\sigma$, the disc's 1-D vertical velocity dispersion and $\Omega$, its angular velocity, while $Q$ is related with the instability of the vertical kinetic to gravitational balance to the excess of centrifugal angular velocity]{Toomre1964}, fragment it into clumps, which largest can reach mass ratios with the disc of a few percent and 1 kpc radii \citep[e.g.,][]{Krumholz2010}. Typical clumps, in galaxies with baryon mass $10^{10}$--$10^{11} M_{\odot}$, range within $\simeq 10^8$--$10^9 M_{\odot}$ \citep[see][]{Agertz2009,Ceverino2010,Ceverino2012}, are rotationally supported, in Jeans equilibrium \cite{Ceverino2010} and stable for long periods ($\simeq 2 \times 10^8$ Myr).
    \item As in local star-generating systems and following the Kennicutt-Schmidt law, a few percents of those clumps' gas converts to stars \cite{Krumholz2010}, while DF migrates them to the DM halo (galactic) center.
    \item The first collapse phase's AC compressed baryons \citep[][and e.g., $10^9M_{\odot}$  galaxy at $z \simeq 5$ in our Fig.~\ref{fig:DFevolDen}]{blumenthal,gnedin}, reenforcing the DM cusp.
    \item The previous effect is countered by DF-induced migration of clumps to galaxy center, as DF transfers energy and angular momentum from baryons to DM, heating up the cusp into a core.
    \item SNF only intervenes later (e.g., around $z=2$), when SN explosions expel gas, decreasing surrounding stellar density. Moreover, such feedback destroys the smallest gas clumps as soon as a small fraction of their mass turns to stars.\footnote{Star formation is not an efficient process.}
\end{itemize}  

Although sharing some common features (e.g., clumps to DM energy transfer via gravitational interaction), leading some to consider SNFF and DFBC as the same idea implemented differently, they fundamentally differ on two crucial points:
\begin{enumerate}
    \item their respective profile flattening epochs differ markedly: while pure SNFF DM profile does not depart noticeably from NFW at $z \simeq 3$ \citep{Pontzen2011,Maccio2012,Onorbe2015}, DFBC flattening initiate at higher redshifts ($z<5$, see Fig.~\ref{fig:DFevolDen});
    \item their clumps moving energy sources diverge in nature: in Refs.~\citep{Mashchenko2006,Mashchenko2008}'s definition, while clumps motion is driven, in the SNFF model, by SN explosions energy, the DFBC views them to just ``passively'' infall to the halo centre.
\end{enumerate}
The two competing scenarios predictions for the central density slope vs. stellar mass and vs. circular velocity relationships were confronted in Refs.~\citep{DPPace2016,DelPopolo:2016skd} with a series of high resolution data predictions from the Milky Way, Sculptor, Fornax, LITTLE THINGS \citep{Oh2015}, THINGS dwarfs \citep{Oh2011a,oh} and Refs.~\citep{Adams2014,Simon2005}. Such confrontation showed slightly better performance from the DFBC model compared with the SNFF scenario, as the former predicts core emergence at smaller stellar masses than $10^6 M_{\odot}$, limiting the latter. Although very small dwarfs ($M_* \le 10^4 M_{\odot}$) cannot produce cores in DFBC, this nevertheless agrees with results from Ref.~\cite{Weidner:2013pnb}.

Both competing model face the same caveats in their resolution of the CC problem: galaxies containing a bulge have been demonstrably able to reconstitute their cusp \citep{DelPopolo2014a}. This effect can even been found in dwarf galaxies \citep{Laporte2015a}.

\subsection{Mass dependent DM 
density profiles}\label{sub:MassDepDen}

Assuming that baryons are indeed responsible for some cored DM haloes, while allowing some others to be cuspy, and using the results of SPH simulations, the SNFF or DFBC models, some authors synthesised the diversity of DM density profiles and generalized the usual density profiles, giving the mass dependent empirical law that would solve the CC problem.
For example, \cite{Lazar2020} generalized the Einasto profile into a mass dependent parameterised profile, encompassing both cuspy and cored cases, while \citep{Freundlich2020} obtained such mass dependent profile by modifying the \citet{Zhao1996} profile.
In the following we will give the details concerning the works of \cite{DiCintio2014b}, and \citep{Freundlich2020}.

In previous discussions, we 
saw that the DM 
density profile is well approximated by the NFW law, and even better fitted 
by the Einasto profile. Those 
two 
profiles are obtained by DM-only simulations. There, the hydrodynamical processes
, fundamental in the determination of the structure of the density profile, are neglected. For example AC 
\cite{Blumenthal1986,Gnedin2004} produced by gas cooling, which 
strengthens cusps, is not taken into account by N-body, DM-only simulations. At the same time, the presence of baryons can expand haloes. As 
discussed in Secs.~\ref{sub:SNsolCC} and \ref{sub:DFclumps}
, stellar feedback and 
DFBC 
can produce a halo expansion and a flattening of the inner density profile. As 
discussed there, \cite{DiCintio2014} showed how the inner slope is modified by SNF
. The results of this paper were used to obtain a DM 
halo whose slope depends from the ratio between stellar  and DM masses
, $\frac{M_*}{M_{\rm halo}}$.

The generalised NFW profile (gNFW) is a specific form of a law depending on three parameters $(\alpha,\beta,\gamma)$:  

\begin{equation}
\rho(r)=\frac{\rho_s}{\left(\frac{r}{r_s}\right)^{\gamma}\left[1 + \left(\frac{r}{r_s}\right)^{\alpha}\right] ^{(\beta-\gamma)/\alpha}}
\label{eq:gNFW}
\end{equation}
being $r_s$ the scale radius and $\rho_s$ the scale density.
$-\gamma$ and $-\beta$, are the logarithmic slopes of the  inner and
outer regions 
respectively, while the sharpness of the transition from inner to outer region is specified by  $\alpha$
. The peculiar case $(\alpha,\beta,\gamma)=(1,3,1)$ corresponds to 
the NFW profile. The goal of \cite{DiCintio2014b} was that of making the profile given in Eq. (\ref{eq:gNFW}) mass dependent, in the sense that its three parameters would depend on the ratio 
$\frac{M_*}{M_{\rm halo}}$, using haloes obtained, in \cite{DiCintio2014} from 
SPH simulations. The density profile of the haloes were fitted, 
using Eq. (\ref{eq:gNFW}).
The behavior of the three parameters was captured using fitting laws as a function of $\frac{M_*}{M_{\rm halo}}$. The outer slope $\beta$ was modeled with 
a parabola
. In the cases of the inner slope, $\gamma$, and of the transition
parameter, $\alpha$, 
double power law model were used
. 

\begin{figure}
\hspace{-0.4cm}
\begin{center}
\includegraphics[width=3.7in]{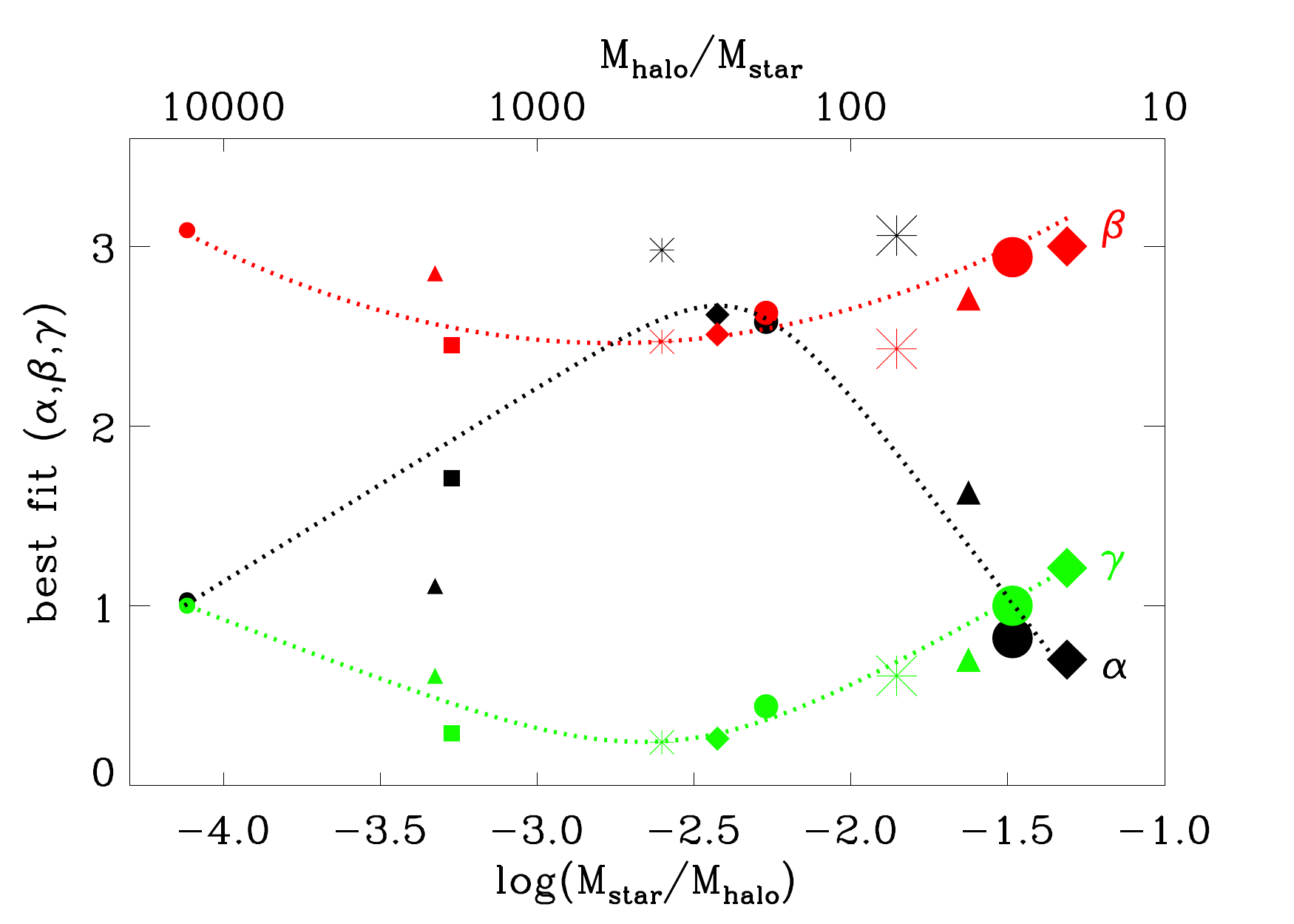}
\end{center}
\caption{Best fits to the generalised NFW parameters $\gamma$
  (green), $\beta$ (red), and $\alpha$ (black) 
  plotted in terms of $\frac{M_*}{M_{\rm halo}}$. The symbols mark various
  SPH simulated galaxies 
  \cite[Figure reproduced from][Figs.~1]{DiCintio2014b}.}
\label{fig:initial_param}
\end{figure}

 The best fits for these parameters to the values for SPH galaxies are shown as dotted lines in
Fig.
~\ref{fig:initial_param}. The functional forms of the fits are:

\begin{equation}
\begin{aligned}
&\alpha= 2.94 - \log_{10}[(10^{X+2.33})^{-1.08}  +  (10^{X+2.33})^{2.29}]\\
&\beta=4.23+1.34X+0.26X^2\\
&\gamma= -0.06 + \log_{10}[(10^{X+2.56})^{-0.68}  +  (10^{X+2.56})]
\end{aligned}
\label{eq:abg}
\end{equation}

\noindent where $X=\log_{10}(\frac{M_*}{M_{\rm halo}})$.

Given 
the stellar-to-halo mass ratio of a galaxy, by means of Eqs.~(\ref{eq:abg}) and 
(\ref{eq:gNFW}), it is possible to compute the entire DM 
profiles.

\begin{figure}
\hspace{-0.4cm}
\begin{center}
\includegraphics[width=3.6in,height=6cm]{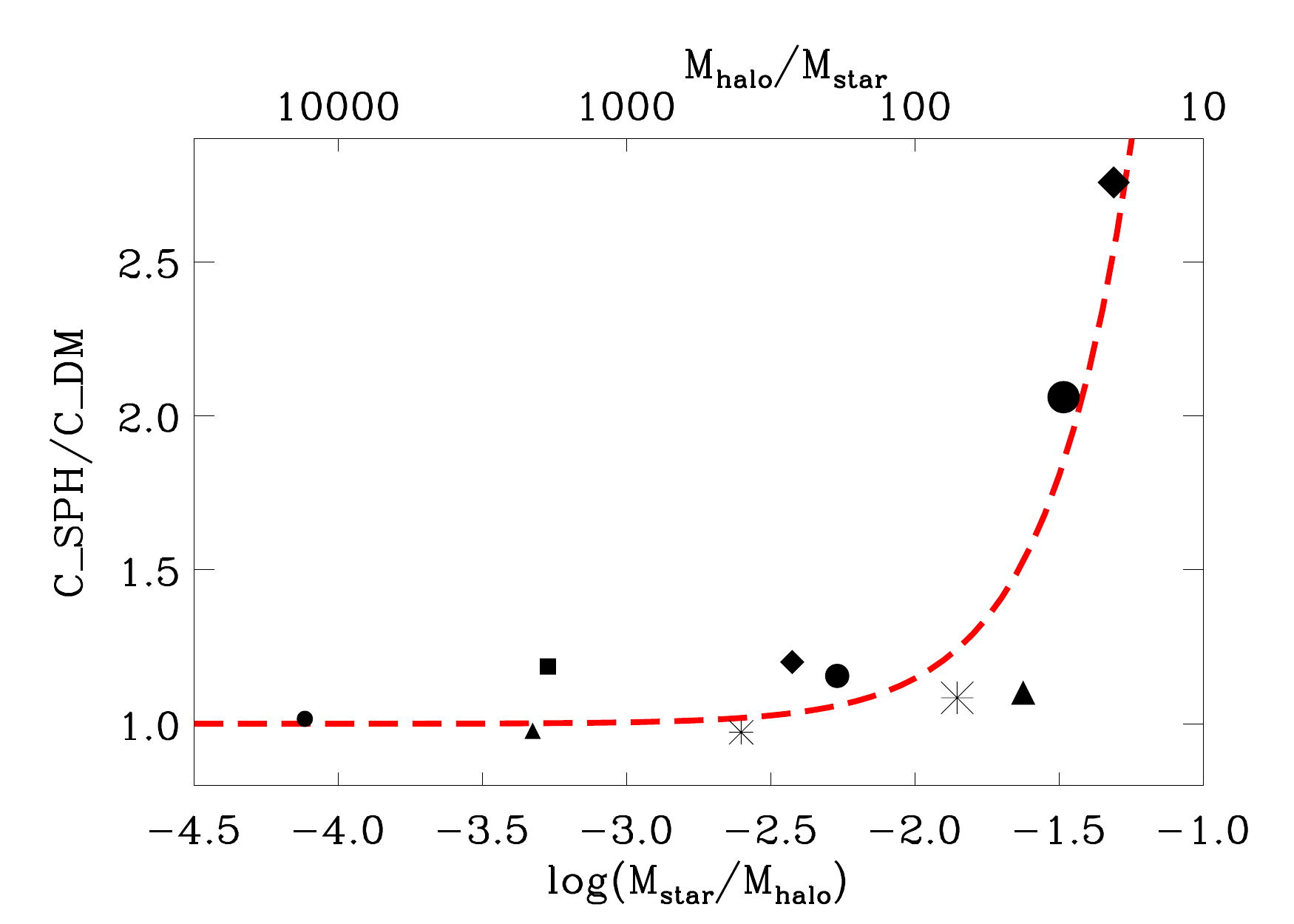}
\end{center}
\caption{Ratio, for a given halo, 
between concentration parameters of its models, in the
  SPH simulation and in the DM-only simulation,  as a
  function of $\frac{M_*}{M_{\rm halo}}$. The best fit to $C_{SPH}/C_{DM}$ is represented by the dashed red
  line \cite[Figure reproduced from][Figs.~4]{DiCintio2014b}.}
\label{fig:conc_vs_magic}
\end{figure}

\cite{DiCintio2014b} also showed that the concentration parameter is dependent on $\frac{M_*}{M_{\rm halo}}$. The dependence of the concentration parameter on $\frac{M_*}{M_{\rm halo}}$ is shown in Fig.~\ref{fig:conc_vs_magic}.  Each
simulation is represented by its symbol and size as described in \cite[Table 1]{DiCintio2014b}
.  The dependence of the ratio between concentration parameters obtained from SPH and pure DM N-body simulations, $C_{SPH}/C_{DM}$, on $\frac{M_*}{M_{\rm halo}}$ is nearly
exponential. Its 
best fit is:
\begin{equation}
 \frac{C_{SPH}}{C_{DM}}=1.0 + 0.00003e^{3.4X_+}
 \label{concentration}
\end{equation}
where $X_+=\log_{10}(\frac{M_*}{M_{\rm halo}}) + 4.5$.

$C_{DM}$ has been derived fitting a NFW profile to the DM-only version of each galaxy, while $C_{SPH}$ has been obtained from hydrodynamical runs where their galaxies DM halo model profile had their scale radius correspondence converted between $r_s$ into $r_{-2}$. 

A comparison of the model with simulated data of the density profiles, and rotation velocities gives a good agreement.

The mass dependent density profile idea from Ref.~\citep{DiCintio2014b} was applied by several other authors. As mentioned above, it was applied by \citet{Lazar2020} to the Einasto profile and by \citet{Freundlich2020} to the so-called Dekel-Zhao density profile, a peculiar case of Eq.~(\ref{eq:gNFW}). We now focus on another popular profile, similar in performance to the Einasto profile, the modified NFW model, adopted by \citet{Freundlich2020} in the form from \citet{Zhao1996}
\begin{equation}
\label{eq:rho_abc}
\rho(r) = \frac{\rho_c}{x^a (1+x^{1/b})^{b(g-a)}}
\end{equation}
where the radial dependence is encoded in $x=r/r_c$, with respect to the characteristic radius $r_c$, and the density scale is set by the typical density $\rho_c$. It is equivalent to the generalised NFW profile, noting the parameters $\alpha=1/b$, $\beta=g$ and $\gamma=a$. Note that, for any natural numbers $n$ and $k$, setting $g=3+k/n$ and $b=n$, it is possible to integrate the mass and obtain the corresponding equilibrium velocity dispersion and gravitational potential analytically. Some choices of integers actually yield very good DM simulations fits, and are even able to accommodate baryons' presence, as well as cusps or core. 

The profile referred to as the Dekel-Zhao (DZ) profile, found for $k=1$ and $n=2$, which corresponds in Eq.~(\ref{eq:rho_abc}) to $b=2$ and $g=3.5$, and leaves two free parameters, $a$ and a concentration parameter $c$, have been found, using the NIHAO suite of simulations 
\citep{Wang2015}, particularly efficient at representing cored profiles \citep[but also cuspy profiles]{Dekel2017}. 

Following Eq.~\eqref{eq:rho_abc},  it takes the form
\begin{equation}
\label{eq:rho32}
\rho(r)  =  \frac{\rho_c}{x^a (1+x^{1/2})^{2(3.5-a)}} 
\end{equation}
using $x=r/r_c$ and defining the concentration $c=R_{\rm vir}/r_c$ from the virial equilibirum radius of the spherical mass distribution $R_{\rm vir}$, where the average density at that radius computes as $\overline{\rho_{\rm vir}}=3M_{\rm vir}/4\pi R_{\rm vir}^3$, while the shape factor $\mu = c^{a-3} (1+c^{1/2})^{2(3-a)}$ is involved in the profile mass integration, yielding the average critical density $\overline{\rho_c}= c^3 \mu \overline{\rho_{\rm vir}}$, and thus relating the density scale to the shape parameters, $c$ and the central slope $a$, as $\rho_c=(1-a/3)\overline{\rho_c}$.

As measurements are made at fixed given resolution, the central slope is actually an integrated slope with respect to the actual profile. \citet{Freundlich2020} explicits, for a fixed resolution $r_1$, an inner logarithmic slope
\begin{equation}
\label{eq:s1_23}
s_1=\frac{a+3.5 c^{1/2}(r_1/R_{\rm vir})^{1/2}}{1+c^{1/2}(r_1/R_{\rm vir})^{1/2}}, 
\end{equation}
and the corresponding averaged concentration parameter
\begin{equation}
\label{eq:c2_23}
c_{2}
=c\left(\frac{1.5}{2-a}\right)^2. 
\end{equation}

The physically motivated restrictions to positive density and inner slope result in the conditions $a\leq 3$ and $a+3.5c^{1/2}(r_1/R_{vir})^{1/2}\geq 0$, respectively. One can reobtain the intrinsic parameters $(a, c)$, employed in analytic expressions, from the observed $(s_{\rm 1}, c_{\rm 2})$, usually bestowed to numerical tests, by inverting Eqs.~\eqref{eq:s1_23} and \eqref{eq:c2_23}, as
\begin{equation}
\label{eq:a(s1,c2)}
a = \frac{1.5 s_1 -2 \left( 3.5-s_1\right)\left(r_1/R_{vir}\right)^{1/2}c_{2}^{1/2} }{1.5 - \left(3.5-s_1\right)\left(r_1/R_{vir}\right)^{1/2}c_{2}^{1/2} }
\end{equation}
and
\begin{equation}
\label{eq:c(s1,c2)}
c= \left( \frac{s_1-2}{\left(3.5-s_1\right)\left(r_1/R_{vir} \right)^{1/2} -1.5 c_{2}^{-1/2}} \right)^2. 
\end{equation}
\citet[Eq. (18), Eq. (21), and Section 2.3, respectively]{Freundlich2020} provided analytic expressions for the DZ profile's gravitational potential, velocity dispersion and lensing properties, as well as, using some NiHAO SPH simulations suite's profiles (see their Fig. 3), compared the DZ, gNFW and Einasto profiles through fits, assuming their parameters free, and finally  finding the DZ profile performs significantly better than Einasto's and marginally better than the gNFW on NiHAO. 

In this context, the mass dependent, DZ-based profile from \citet{Freundlich2020}, was constructed, as in \citet{DiCintio2014b}, from the NIHAO simulations suite \citep{Wang2015}, to set the DZ parameters functional forms. SPH simulations were fitted by \citet{DiCintio2014} with a similar profile to Eq. (\ref{eq:rho_abc}), to set the free parameters forms as a function of $M_{\rm star}/M_{\rm vir}$, as did \citet{Lazar2020} with the Einasto profile and \citet{Freundlich2020} with the DZ profile. 
However, the mass dependence introduced, similarly to Eqs.~(\ref{eq:abg}) by \citet{DiCintio2014b}, to improve galaxies and clusters density profiles fits, such as in the modified Einasto profile from \citet{Lazar2020} or in the modified NFW from \cite{Read2016}, does not allow to keep any analytic form for the mass, as detected through lensing, velocity dispersion or gravitational potential. On the contrary, the DZ profile keeps that possibility intact.

This is the advantage of the DZ profile fit by \citet{Freundlich2020}, applied to simulated NIHAO haloes' density profile logarithms, optimising the parameters, spacing the $N \simeq 100$ grid profile radii $r$ in the range 0.01 $R_{\rm vir}$-$R_{\rm vir}$, through a least-square minimization. Setting the virial mass $M_{\rm vir}$, the profile can set the scale by extracting the corresponding $R_{\rm vir}$, leaving the remaining free parameters $a$ and $c$. The star distribution can be samely fitted to get the star mass at virial radius $M_{\rm star}$ and obtain the corresponding $M_{\rm star}/M_{\rm vir}$. The theoretical slope and concentration are related to the integrated (observed) ones through Eqs.~\eqref{eq:s1_23} and \eqref{eq:c2_23}, and Eqs.~\eqref{eq:a(s1,c2)} and \eqref{eq:c(s1,c2)}, respectively.

\begin{figure*}
\begin{centering}
\includegraphics[width=0.8\columnwidth]{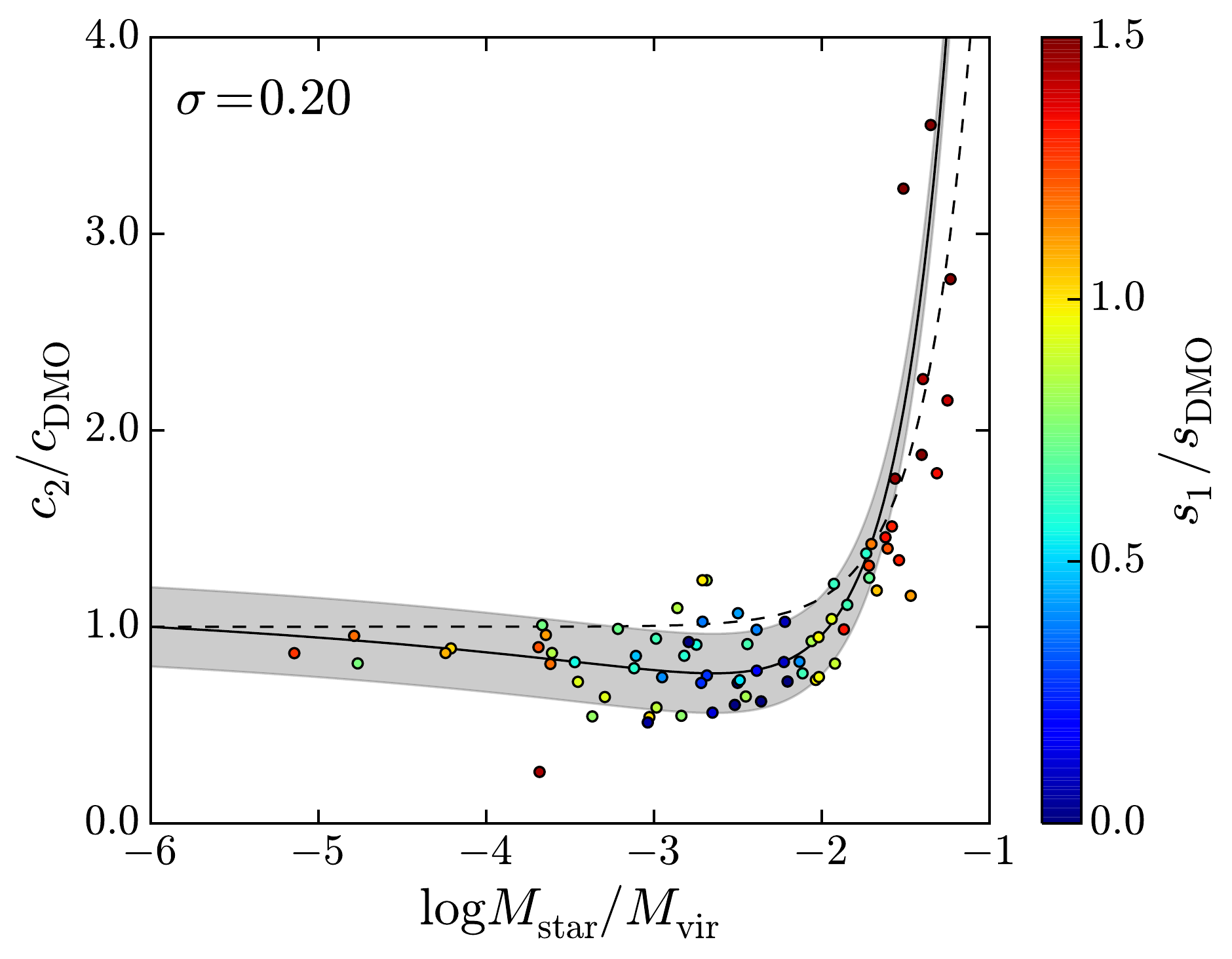}
\includegraphics[width=0.8\columnwidth]{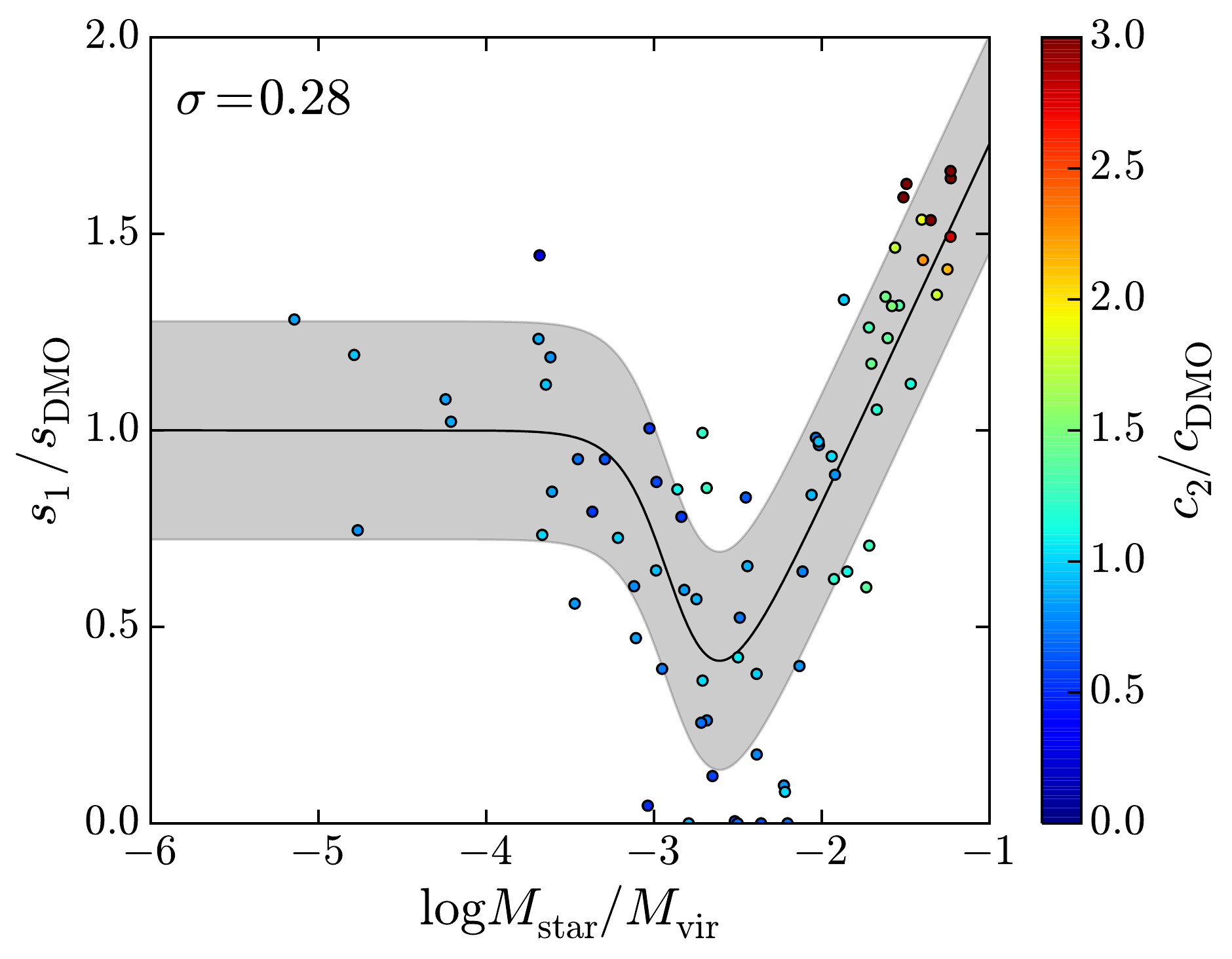}
\par\end{centering}
\caption{\label{fig:NSWD_branches_y01}
Ratios, for integrated slopes and concentrations $s_1$ (lower panel)
and $c_2$ (upper panel), between the  DZ model parameters and their DM-only NFW profiles counterparts, as a function of mass ratios $M_{\rm star}/M_{\rm vir}$.  The NFW integrated parameters, $s_{DMO}$ and $c_{DMO}$, were obtained from the \citet{Dutton2014}
relation \cite[Eqs. 47 and 48 of][and their best fits, Eqs. (45) and (49) on their Fig.~9, here marked as plain black lines]{Freundlich2020}. The best fit residual rms $\sigma$ is shown in gray. The dashed black line in the $c_2/c_{DMO}$ panel represents the best-fit function from \citet{DiCintio2014b}. Color coding, indicated on the right side in each panel, for each halo compounds the information of the other panel.
}
\end{figure*}

This fitting procedure allowed to obtain, for each NIHAO haloes, their $s_1$ and $c_2$ as a function of $M_{\rm star}/M_{\rm vir}$, as seen  \citet[][Fig. 8, black dots]{Freundlich2020}. 

The obtained $s_1$ and $c_2$ behaviour were synthesised in two parameterised functions of $x=\frac{M_*}{M_{\rm halo}}$, where $x_0$, $s^\prime$, $s^{\prime\prime}$, and $\nu$ are ajustable and which used values were given in \citet[Table 1]{Freundlich2020}
\begin{equation}
\label{eq:s1(x)}
s_1(x) = \frac{s^\prime}{1+\left(\frac{x}{x_0}\right)^\nu} + s^{\prime\prime} \log \left(1+\left(\frac{x}{x_0}\right)^\nu\right),
\end{equation}
while the parameter $c^\prime$ is introduced for
\begin{equation}
\label{eq:exp_function_c0}
c_2(x)= c^\prime \left(1+\left(\frac{x}{x_0} \right)^\nu\right).
\end{equation}

The resulting mass-dependent profile presents the advantages, over the 
\citet{DiCintio2014b} approach, to require less free parameters as well as to allow integration into analytic expressions for lensing properties, dispersion velocity and gravitational potential.


\section{Cosmological Solutions to the CC Problem}\label{sec:CosmoSol}

The CC problem, together with the other small scale problems encountered by the $\Lambda$CDM model, open the question of a more general failure of the CDM's paradigm.  This justifies models in which the nature of DM or gravity itself are modified. Such models encountered various degrees of success in their already widely checked instances. We will here focus on alternate models of DM. Several approaches have been used and we will detail the two main avenues, before discussing other possibilities.


\subsection{Warm DM}

Warm DM (WDM) presents the simplest departure from CDM, endowing DM with a small velocity dispersion ($\sigma\simeq 100$ m/s nowadays), also linked, for thermally produced DM, to a small DM relic mass \citep[][]{Colin2000,som_dol}. Since velocity dispersion is expected to decrease with time, it should be more important in the past and thus smear small scale structure formation. Such smear induces similar DM ``heating'' as the baryonic solutions of Sec.~\ref{sub:BsolCC}: in this case the WDM particles higher velocity than for CDM smears the haloes to flatter profiles as well as produce fewer low mass haloes. These effects of WDM on structure formation have been copiously simulated \citep[e.g.,][]{PolisenkyRicotti2011,Lovell,Maccio2012,Angulo}.

Despite solving the CC problem for haloes with their corresponding scales on a case by case basis by tuning its particle mass, WDM does not manage to solve it with one mass for all galaxies, nor in the entire CC problem mass range \citep{kuzio2010}. As shown in Ref.~\cite{Wu:2014eva}, neither WDM or pure collisionless CDM models can match observed disk galaxies rotation curves, while they can agree with hydrodynamical simulations including baryons (see Sec.~\ref{sec:ccp}'s discussion).

\begin{figure}[t]
\centering
\resizebox{7cm}{!}{\includegraphics{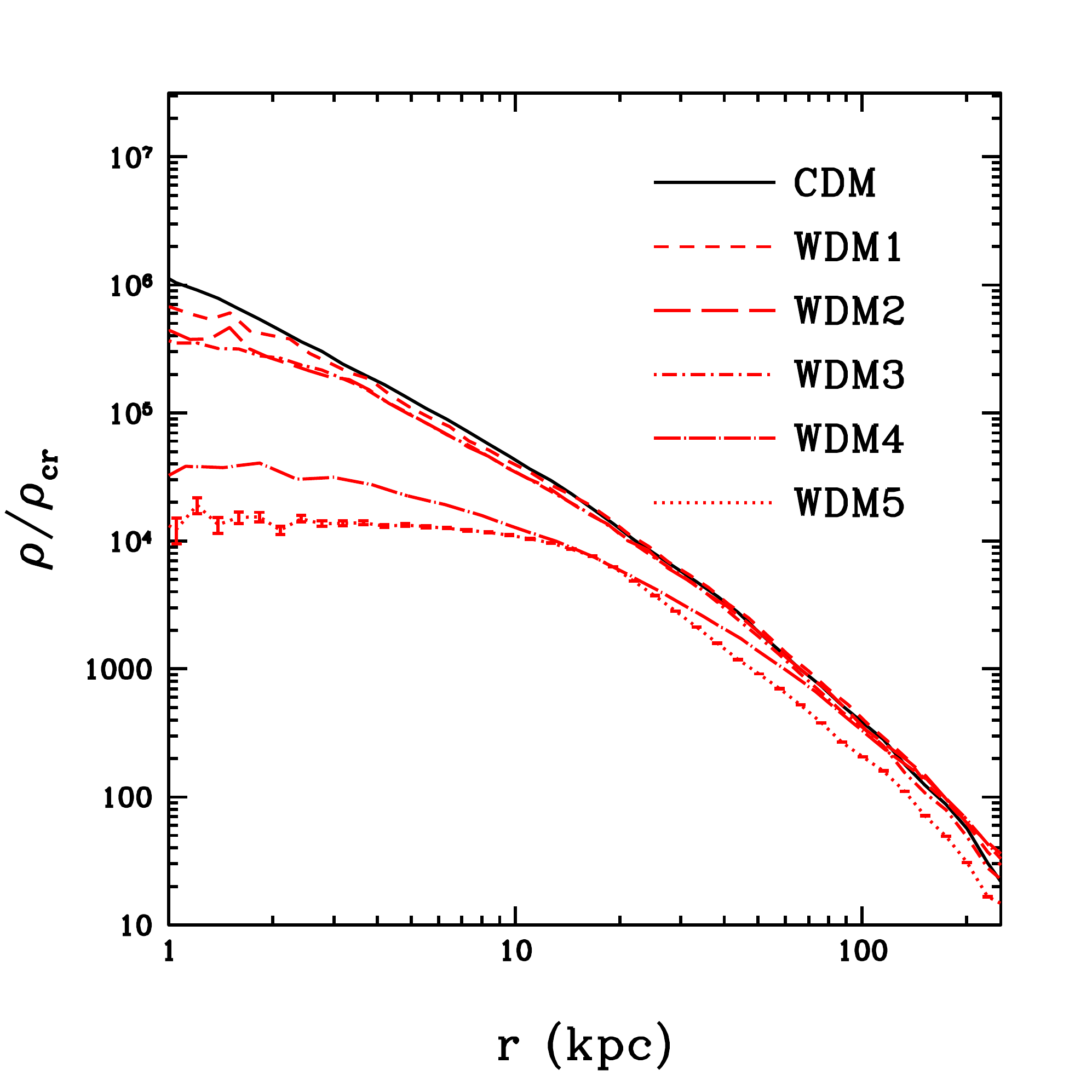}}
\resizebox{6.6cm}{!}{\includegraphics{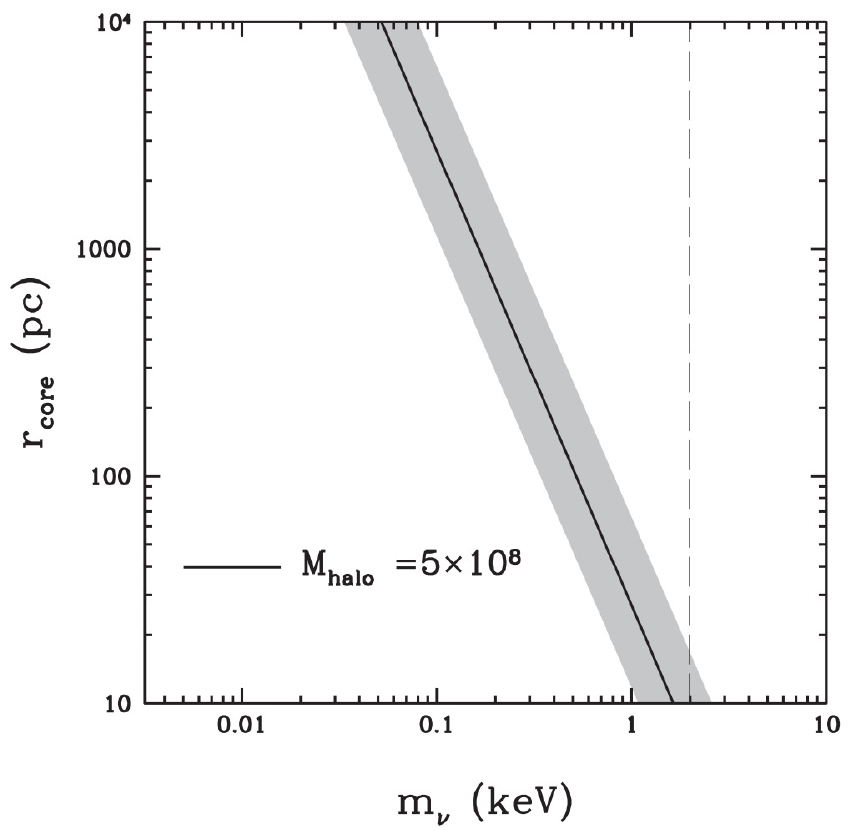}}
\caption{
WDM effect on the CC problem, reproduced from \citep[][Figs.~2 and 8]{Maccio2012b}. The WDM core formation is illustrated against the CDM density profile (solid black line) by showing profiles from a range of five WDM particle masses, from $m=2 keV$ (WDM1, as per the legend) to 0.05 keV (WDM5), in the left panel. The right panel synthesises the WDM core radius obtained for its given particle mass with other constraints: while the solid black line marks the core size dependence at fixed halo mass, the shaded area limits the plane to the cosmological constraints for $0.15 < \Omega_m < 0.6$ and the vertical dashed line sets the upper limit from large scale observations.
 }\label{fig:WDM}
\end{figure}

In addition, Ref.~\citep[][for a $m=2$ keV thermal relic]{PolisenkyRicotti2011}, and several other authors \citep[e.g.,][]{Dalal,Fadely,Fadely1}, shown that, for instance, the strong-lensing subhalo fraction is too high compared with WDM produced subhaloes. Furthermore, WDM thermal mass displays tensions between expected core size (1 kpc core corresponding to 0.1 keV) and large scale structure \citep[$m \simeq$ 1--2 keV leads to 10--20 pc cores, see Fig.~\ref{fig:WDM} and Ref.][]{Maccio2012b}. In conclusion, because its power spectrum falls off too sharply, WDM does  not improve on CDM \citep{Schneider}. Finally, since WDM structure formation is also shown to modify the Lyman-$\alpha$ forest \citep{Nayara}, the situation is even worse, as WDM solution to the CC problem in our vicinity cannot, at the same time, be consistent with observed high redshift Lyman-$\alpha$ forest.

\subsection{Self-Interacting DM}
The second simplest variation away from CDM, coined Self-interacting DM \citep[hereafter SIDM, ][]{sperg_ste}, consists in granting to DM a self-interaction cross-section, within the same magnitude of nucleon-nucleon interactions ($\simeq$(m/g)$^{-1}$cm$^2$).\footnote{Note that 1 cm$^2$/g $\simeq$ 1 barn/GeV, so multiplying by the DM particle mass yields the  cross-section.} The resulting elastic scattering in regions of highest density (inner galactic regions) redistributes energy and angular momentum so that tri-axiality is reduces and a Burkert profile core forms \citep{burkert}. 

\begin{figure}[t]
\centering
\resizebox{11cm}{!}{\includegraphics{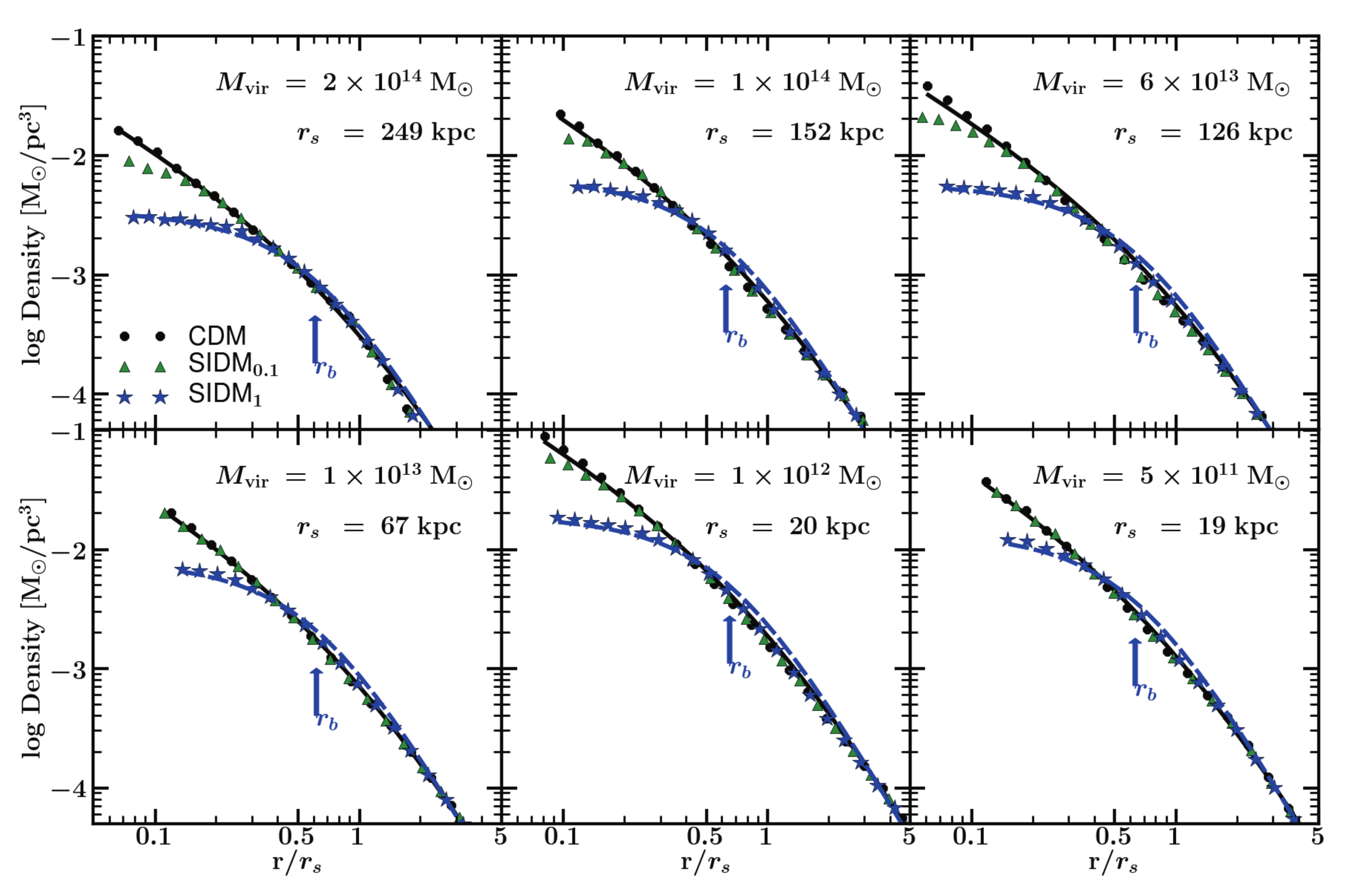}}
\caption{SIDM simulations compared with CDM, indicated by black dots, in density profiles, reproducing a figure from Ref.~\citep[][Fig.~4]{rocha}. Two SIDM models, designated in the legend as $SIDM_{1}$ and $SIDM_{0.1}$, are shown, using DM particle masses $\sigma/m=1$, marked with blue stars, and $0.1$, characterised with green triangles, respectively. They are ovelayed with the NFW, plotted as a black line, and the Burkert, traced in blue, density profiles. The core radii of Burkert's profiles are located with arrows.
}\label{fig:SIDM}
\end{figure}

The CC problem has been claimed to be solved in dwarfs, MW-sized galaxies, and galaxy clusters by SIDM from some cosmological simulations~\citep{Newman2013a,newman2,rocha,peter} consistent with merging galaxy clusters observations \citep[][0.1--0.5 cm$^2$/g cross sections]{clowe,randall,dawson}. Fig.~\ref{fig:SIDM} presents simulations of haloes from galaxies to clusters masses, with $\sigma/m$ set within two values, in the case of Ref.~\citep{rocha}. Although SIDM appears to solve the CC problem, since its cored subhaloes are more sensitive to disruption and tidal stripping than CDM's, and to improve on WDM \citep{rocha,peter}, as enough subhaloes survive, a different point of view still remains \citep[][]{kuzio2010}.

The SIDM's attraction does not entirely resides in its structure formation effects. As ``hidden sector'' particle models naturally produce them  \citep[e.g.][]{sperg_ste,Yoshida2000,Dave2001}, it also retains appeal from the particle theory point of view.

\subsection{Other DM models}
Other proposals to change from the CDM model are more or less loose declinations altering around the SIDM concept. Here is a list of the main variations:
\begin{description}
\item[RDM] switching interaction to negative scattering results in repulsive DM \citep[RDM,][]{goodman}.
\end{description}
The next two variations from SIDM are very commonly related to indirect DM detection:
\begin{description}
\item[SADM] if self-interaction produces a DM particles annihilation, the result is coined Self-Annihilating DM \citep[SADM,][proposing cross section-velocity $\sigma v \simeq 10^{-29}$ (m/GeV) cm$^2$]{kap}. 
Such annihilation decreases the dense regions' particle numbers, in particular in the halo's centre, reducing central gravity, therefore allowing to expand central particles' orbits and thus to flatten the central profile. At the same time, annihilation results in radiation emissions that are possibly detectable.

\item[DDM] Alternately, self-interaction can result in DM decay into relativistic particles, designated as Decaying DM \citep[DDM,][]{cen1}. Their gravitational effect on structures is similar to SADM, as they equally deplete galaxies' central density, since the relativistic particles escape away, while larger scales structures behaviour remain similar to CDM. Similarly, the relativistic decay products produce radiations that can also be detected.
\end{description}
The last main variations are more loosely related to SIDM:
\begin{description}
\item[BCDM] Superfluid behaviour of non-relativistic, massive boson condensates in haloes centre can also result in smoothing down their profiles from cusp to cored \citep{harko}. Recent structure formation simulations \citep{schive} demonstrated that the Bose condensate DM (BCDM) small scale gravity vs. uncertainty principle opposition reduced substructures and flattened the density profile at those scales, while producing larger scale structures indistinguishable from CDM outputs.
\item[SFDM] As scalar field condensation also forms a Bose condensate, similar flat galaxies inner profiles were obtained from such implementation of the BCDM \citep{robles}, that was called Scalar Field Dark Matter (SFDM)
\item[FDM] Further implementation of the gravity vs. uncertainty principle exploited the wave-particle duality to obtain Fuzzy DM \citep[FDM,][]{hu}, using galactic core sized Compton wavelength, ultra-light ($m \simeq 10^{-22}$ eV) scalar particles. As they cannot be ``squeezed'' below their Compton wavelength, they also develop flatter profiles and form less substructures.
\end{description}

\section{Conclusions}\label{sec:concl}

Since the first assessment by \"Opik of the amount of matter surrounding our solar system through vertical star motion with respect to the plane of the ecliptic more than a century ago \cite{Opik}, current data indicates a dominant amount of matter should be in the form of DM. However, such dynamics and geometric data has still not been confirmed, to date, by direct or indirect DM particles detection \cite{dp13,dp14a}.

Grasping the essence of DM from a gravitational point of view should privilege the study of galaxy satellites, dwarf galaxies and other small scale structures, dominated by DM, all the more as they present significant departures from the $\Lambda$CDM model predictions. In  this review, we focused on the departure constituted by the CC problem, formulated more than twenty years ago. Although we now have some competing explanations, there is no concensus on the CC problem's cause.

In this review, we have discussed the focus on baryonic effects on small scale structures. We recalled that the DFBC model can explain cores in $M >10^5 M_{\odot}$ dwarfs from baryon clumps interactions with DM, while cusps form in MW-type galaxies \cite{DPPace2016}. Although the SNFF model gets similar results, its prediction cover structures in a more restricted range $M_*>10^6 M_{\odot}$ \cite{Governato2012}. Note that both models, by depleting the DM density in central halo regions, also explain other small scale problems such as the TBTF or the subhaloes abundance problems. However, baryonic solutions to the CC and other small scale problems are still debated, as some issues remain, such as with the SNFF in classical MW dwarfs \cite{Penarrubia2012,GarrisonKimmel2013}, or the possible future discovery of $M_*<10^6 M_{\odot}$ dwarfs with flat inner profiles which would drive the SNFF model to conclude that DM is actually not cold\footnote{Note that such conclusion is not true for the DFBC model.}, or even that no baryonic physics solutions clearly appear for the TBTF problem in isolated galaxies. Escaping this debate would entail understanding how the SIDM model or one its variants, as we have discussed, could solve the CC problem. Future surveys, including measurements of star velocities, would anyhow be required to further our understanding of the CC problem and other small scale issues.

From an observational point of view, as discriminating cores from cusps in dSphs remains delicate, the discussion on their structure persists \citep[see the introduction and, e.g.,][]{Strigari2014}. Although the Subaru Hyper-Supreme-Camera~\cite{Takada} and GAIA \cite{Bruijne} were foreseen as providing possible resolution for the question, their abilities would not exceed larger dwarf galaxies, of the size of Sagittarius \cite{Richardson2014}. Based on Jeans equations, their determination of density profile using stars line of sight velocity and 2D projection radius are plagued with degeneracy with the anisotropy parameter. Notwithstanding the challenges GAIA faces from it \cite{Richardson2014}, improvements to measure dwarfs's stars proper motions of the \cite{strigari}, that yield  density slope at half-light radius, were recommended, that focused on only one component out of the three velocities and on just two components out of three positions \citep[see][]{Battaglia2013}. However successful these efforts may be, understanding dwarfs inner structure will be crucial. In contrast, the difficulties of the SNFF tensions do not appear fundamental: baryons, through the more efficient DFBC model, can still turn cusps into cores sooner than gas turns into stars. Consequently, should cored dwarfs with $M_* < 10^6 M_{\odot}$ be discovered, despite their inconsistency with the SNFF, provided $M_* > 10^5 M_{\odot}$, the $\Lambda$CDM model correct with baryonic effects of the DFBC model would remain consistent with observations.

Beyond the CC problem and the other documented issues in small scales of the $\Lambda$CDM model, further testing could involve
\begin{itemize}
    \item checking the predicted large subhaloes number within galaxies' virial radius, barely agreeing with anomalies in the gravitational lensing fluxes \cite{MetcalfAmara};
    \item measuring subhaloes perturbations of the MW's cold tidal streams \citep[see][]{Carlberg}.
\end{itemize}

Finally, the nature and existence of DM would find a definitive solution in detection of DM particles, whether in colliders, direct or indirect means. Despite the hopes raised by a di-photon excess decay at 750 GeV at 3.9 $\sigma$ significance in the LHC, as it was later ruled out as a statistical fluctuation, no evidence of super-symmetry (SUSY),  that would allow WIMPS particles DM, have so far emerged and collider detection of DM is yet elusive. Although controversial, the annual modulation claimed in DAMA/LIBRA/CoGeNT remains the only possible direct DM particle detection. To date, no incontrovertible direct, indirect or collider evidence for the nature or existence of DM has been announced.

\vspace{6pt}


\acknowledgments{MLeD acknowledges the financial support by the Lanzhou University starting
fund, the Fundamental Research Funds for the Central Universities
(Grant No. lzujbky-2019-25), National Science Foundation of China (grant No. 12047501) and the 111 Project under Grant No. B20063.}

\authorcontributions{All authors have contributed equally in this work.}

\conflictofinterests{{The authors declare no conflict of interest.}}

\bibliographystyle{mdpi}

\renewcommand\bibname{References}
\bibliography{lite,shortnames,old_MasterBib2,biblioNS,freu}
\end{document}